\documentclass[11pt,a4paper]{article}
\pdfoutput=1
\usepackage{jheppub}
\usepackage{gensymb}
\usepackage{subcaption}
\usepackage{mwe}
\usepackage{amssymb,amsmath}
\usepackage{graphicx}
\usepackage{color}
\usepackage{cancel}
\usepackage{booktabs}
\usepackage[colorlinks=true
,urlcolor=blue
,citecolor=blue
,linkcolor=blue
,pagecolor=blue
,linktocpage=true
,pdfproducer=medialab
]{hyperref}

\usepackage[numbers]{natbib}
\usepackage{notoccite}
\usepackage{comment} 
\usepackage{placeins}
\usepackage{multirow}
\usepackage{setspace}
\usepackage{bigints}
\makeatletter \renewcommand{\@dotsep}{10000} \makeatother

\newcommand{\mgut}{M_{{\rm GUT}}}

\newcommand{\mmess}{M_{{\rm Mess}}}

\newcommand{\cred}[1]{{\bf \color{red} #1}}

\DeclareUnicodeCharacter{2212}{-}
\DeclareUnicodeCharacter{202F}{\,}
\begin{document}

\begin{titlepage}
\pagestyle{empty}

\vspace*{0.2in}
\begin{center}
{\Large \bf Non-Holomorphic Impact on $t-b-\tau$ Yukawa Unification \\ in minimal GMSB}

\vspace{1cm}
{\bf Ali \c{C}i\c{c}i\footnote{E-mail: ali.cici@cern.ch}$^{,a,b}$, B\"{u}\c{s}ra Ni\c{s}\footnote{E-mail:busranis@uludag.edu.tr}$^{,c}$ and Cem Salih $\ddot{\rm U}$n\footnote{E-mail: cemsalihun@uludag.edu.tr}$^{,c}$}
\vspace{0.5cm}

{\small \it  $^{a}$Orhaneli T\"{u}rkan-Sait Y\i lmaz High School, TR16980 Bursa,T$\ddot{u}$rkiye 
\\  $^b$Department of Physics, Faculty of Engineering and Natural Sciences, \\ Bursa Technical University, TR16310, Bursa, T\"{u}rkiye
\\ $^{c}$Department of Physics, Bursa Uluda\~{g} University, TR16059 Bursa, T$\ddot{u}$rkiye}

\end{center}

\vspace{0.5cm}
\begin{abstract}

We study and explore the low scale implications of Yukawa unification in the minimal gauge mediated supersymmetry breaking models. We also assume non-zero non-holomorphic terms, with which the YU solutions can be accommodated in the cases with $\mu > 0$. These results can be considered a direct effect from the non-holomorphic terms, but they also lead to testable low scale implications compatible with YU. We observe abundant solutions consistent with the Higgs boson mass. This constraint leads to heavy strongly interacting supersymmetric particles, while the electroweak sector can be relatively light and they can be probed by several experiments. We find that the chargino can be lighter than 1 TeV and it is degenerate with the lightest neutralino in most of these solutions. Consistent solutions in this region can accommodate charginos as light as about 120 GeV, and they will likely be tested soon by the analyses over the compressed spectra. These solutions are also subjected in the lifetime analyses. Our analyses identify such light charginos as decaying in about $10^{-2}$ ns. Probing such points may need a slight improvement in the sensitivity of current analyses, and one can expect them to be tested very soon. In the same region, the stau is found to be another light supersymmetric particle, and some of the solutions can be inconsistently lighter. We find that it can weigh as light as about 600 GeV consistently, and it can be tested also soon over its lifetime. We summarize and also exemplify our findings with five benchmark scenarios. Furthermore, the benchmark solutions also reveal that the solutions can be tested in heavy Higgs boson searches, which shape the whole Higgs boson spectrum in models.

\end{abstract}
\end{titlepage}

\section{Introduction}
\label{sec:intro}

In this study we continue exploring the impact of the non-holomorphic (NH) terms on the Yukawa unification (YU) within the minimal gauge mediated supersymmetry (SUSY) breaking (mGMSB) models. As a preferred framework, we consider the boundary conditions which cannot accommodate YU when such NH terms are absent. {The results discussed in this study are presented as a completion of a recent study of ours \cite{Cici:2026lty}, in which we discussed the impact of the NH terms on YU in detail within a class of Yukawa unified gravity mediated SUSY breaking models. In contrast to gravity mediated SUSY breaking models, GMSB models can provide a more restricted framework for the NH terms, and one can explore their limits within which they can still yield visible impacts on the low scale implications. Since they are expected to modify the typical mass spectrum compatible with YU, one can identify light electroweakly interacting SUSY particles which can be subject to several experimental analyses.  Because mGMSB models typically accommodate a gravitino as the lightest supersymmetric particle  (LSP) in the mass spectrum, one of the distinct signatures for these models can be probed via the long lived particles \cite{CMS:2021tkn,CMS:2024trg,ATLAS:2025fdm,ATLAS:2026hnb}.}

In the usual treatments, SUSY is assumed to be broken in a hidden sector, and its breaking is transmuted to our sector through some messengers which are allowed to interact with the SUSY partners of the Standard Model (SM) particles. If the messenger fields participate in these interactions through the gauge interactions of SM, the models can be classified in GMSB \cite{Giudice:1998bp,Meade:2008wd,Dimopoulos:1996yq}. Apart from ameliorating the gauge hierarchy problem, these models are also favored by the flavor analyses, since SUSY breaking is realized through the flavor blind gauge interactions. In addition, the trilinear soft SUSY breaking (SSB) terms ($A-$terms) are generated with negligible magnitudes, and it could be another advantage, since the stability of the scalar potential favors low $A-$terms for the third family sfermions \cite{Ellwanger:1999bv}. In addition, from theoretical point of view, the SUSY breaking scale ($M_{{\rm Mess}}$) in the visible sector could be as low as about the electroweak scale; thus, one can recover SUSY even at the scales to which the current collider experiments can be highly sensitive.

Despite such theoretical advantages, models in this class have received the strongest negative impact from the Higgs boson discovery \cite{ATLAS:2012yve,CMS:2012qbp,CMS:2013btf}, especially as a consequence of the small $A-$terms. Despite their smallness, the $A-$terms can be driven to relatively large values through the renormalization group equations (RGEs); however, the solutions with large $A-$terms at the low scale required by the Higgs boson mass can be realized only when the gluino and squarks are quite heavy and/or the SUSY breaking occurs at a high energy scale \cite{Ajaib:2012vc}. When the GMSB models are established in a minimal way, which imposes one multiplet for the messenger fields (say, $\mathbf{5} + \mathbf{\bar{5}}$ of $SU(5)$), the requirement of heavy SUSY spectra also affect the sleptons and all the gauginos.

The results summarized in the paragraph above have been excessively explored in the absence of the NH terms. This is because, if there is no other source to generate such terms, they are induced through the SUSY breaking, and their magnitudes are suppressed by the SUSY breaking scale compared to the SSB masses of SUSY scalars \cite{Bagger:1995ay,Martin:1999hc}. However, they can still be considerable when the SUSY breaking scale is relatively small as can happen in GMSB models. Earlier studies have shown that the low scale implications of the GMSB models can be drastically changed when the contributions from NH terms are taken into account \cite{Chattopadhyay:2017qvh,Ali:2021kxa,Chakraborty:2019wav,Un:2014afa,Rehman:2022ydc,Un:2023wws,Israr:2025cfd,Israr:2024ubp,Rehman:2025djc,Nis:2025fxc}. One of the interesting impacts in these cases can be observed in the Higgs boson mass. The NH terms can significantly contribute to the Higgs boson mass, and they can loosen the impact from the need for heavy SUSY spectra especially in the models with small $A-$terms.

As mentioned at the beginning, we consider models in the mGMSB class in which all of the SSB terms are induced through interactions between the messengers from $\mathbf{5}+\mathbf{\bar{5}}$ representation of $SU(5)$ and the SUSY and Higgs fields. We impose the RGEs running above the SUSY breaking messenger scale ($M_{{\rm Mess}}$) in the presence of the messenger fields to impose gauge unification and YU within this class. In addition, we also consider the NH terms which can be effective after SUSY breaking. The rest of the paper is organized as follows: We first summarize the RGEs above $M_{{\rm Mess}}$ to impose the unification of the gauge and Yukawa couplings in Section \ref{sec:NHYU}. We also briefly discuss the NH terms and their contributions to Yukawa couplings. After we describe the scanning procedure and experimental constraints in our analyses in Section \ref{sec:scan}, we discuss our results and prospects for testing the YU implications at the current and/or future experiments. Finally, we conclude and summarize our findings in Section \ref{sec:conc}.

\section{Yukawa Unification in NH mGMSB}
\label{sec:NHYU}

The YU condition in the mGMSB framework can be explored by supplementing the model with RGEs above $M_{{\rm Mess}}$. The superpotential above $\mmess$ involved the messenger fields as follows \cite{Giudice:1998bp}:

\begin{equation}
W = W_{{\rm MSSM}} + \hat{S}\hat{\Phi}\hat{\bar{\Phi}}~,
\end{equation}
where $\hat{S}$ is an MSSM singlet field which breaks SUSY in the hidden sector through its vacuum expectation values as $\langle \hat{S} \rangle = \langle S \rangle + \theta^{2}\langle F \rangle$, and $W_{{\rm MSSM}}$ represents the usual MSSM superpotential. The scales for the SSB terms can be parametrized with $\Lambda \equiv \langle F \rangle / \langle S \rangle$, which splits the masses of the scalar ($\Phi$) and fermionic ($\Psi_{\Phi}$) messenger fields as 

\begin{equation}
m_{\Phi} = M_{{\rm Mess}}\sqrt{1\pm \dfrac{\Lambda}{M_{{\rm Mess}}}}~; \hspace{0.3cm} m_{\Psi_{\Phi}} = M_{{\rm Mess}}
\label{eq:mess}
\end{equation}

\subsection{Above $\mathbf{M_{{\rm Mess}}}$}
\label{subsec:AMmess}

Since all of the SSB terms including the NH ones are induced below $M_{{\rm Mess}}$, the RGEs between $M_{{\rm Mess}}$ and $\mgut$ involve only the gauge and Yukawa couplings as follows:

\begin{equation}
\frac{d g_a}{dt}=\frac{1}{16\pi^2}\left(\beta^{(1)}_a+g^3_a N_5\right)
        +\frac{1}{(16\pi^2)^2}\left(\beta^{(2)}_a+g^3_a \sum_{b=1}^3 2N_5  B^{(2)}_{ab}g^2_b\right)~,
\label{eq:abRGE}
\end{equation}
where $N_{5}$ is the number of the messenger multiplets and it is set as $N_{5}=1$ in our work. $g_{a}$ with $a=1,2,3$ denote the gauge couplings of $U(1)_{Y}$, $SU(2)_{L}$ and $SU(3)_{C}$, respectively. The $\beta^{(1,2)}_a$ are one- and two-loop MSSM beta functions \cite{Martin:1993yx}, and
\begin{equation}
B^{(2)}_{ab}=
\begin{pmatrix}
7/30 & 9/10 & 16/15 \\
3/10 & 7/2 & 0\\
2/15 & 0 & 17/3
\end{pmatrix}
\label{eq:Bab}
\end{equation}
is the contribution from a single 5-plet. Note that $5$ and $\bar{5}$ give the same contribution to the RGEs. The RGEs for the Yukawa couplings above $M_{{\rm Mess}}$ can be given as follows \cite{Martin:1993yx,Gogoladze:2015tfa}:

\begin{equation}
\setstretch{2.0}
\begin{array}{rl}
\dfrac{d y_t}{dt}&=  \dfrac{1}{16\pi^2}\beta^{(1)}_{t}
                 +\dfrac{1}{(16\pi^2)^2}\left[\beta^{(2)}_t
                     +y_t N_5\left(\dfrac{8}{3}g_3^4+\dfrac{3}{2}g_2^4+\dfrac{13}{30}g_1^4 \right) \right],\\
\dfrac{d y_b}{dt}&=  \dfrac{1}{16\pi^2}\beta^{(1)}_{b}
                 +\dfrac{1}{(16\pi^2)^2}\left[\beta^{(2)}_b
                     +y_b N_5\left(\dfrac{8}{3}g_3^4+\dfrac{3}{2}g_2^4+\dfrac{7}{30}g_1^4 \right)\right],\\
\dfrac{d y_{\tau}}{dt}&=  \dfrac{1}{16\pi^2}\beta^{(1)}_{\tau}
                   +\dfrac{1}{(16\pi^2)^2}\left[\beta^{(2)}_\tau
                     +y_\tau N_5\left(\dfrac{3}{2}g_2^4+\dfrac{9}{10}g_1^4 \right)\right]~.
\end{array}
\label{eq:yuRGE}
\end{equation}

Models in the mGMSB class can be constrained with the YU condition after calculating the Yukawa couplings at $\mgut$ by running the RGEs given above between $M_{{\rm Mess}}$ and $\mgut$. One can utilize the following parameter with a suitable criterion to identify the solutions compatible with YU:

\begin{equation}
R_{tb\tau} = \dfrac{{\rm max}(y_{t},y_{b},y_{\tau})}{{\rm min}(y_{t},y_{b},y_{\tau})}~,
\end{equation}
in which the Yukawa couplings are considered at $\mgut$.

\subsection{Below $\mathbf{M_{{\rm Mess}}}$}
\label{subsec:BMmess}

Once the masses are split between the fields in the messenger supermultiplets as given in Eq.(\ref{eq:mess}), they induce the following SSB Lagrangian involving the MSSM fields through loop levels:

\begin{equation}
\setstretch{2.0}
\begin{array}{ll}
\mathcal{L}^{{\rm MSSM}}_{{\rm soft}} = & -\dfrac{1}{2}\left(M_{1}\tilde{B}\tilde{B} + M_{2}\tilde{W}\tilde{W} + M_{3}\tilde{g}\tilde{g} \right) \\
& - \left(\tilde{Q}^{\dagger}m_{Q}^{2}\tilde{Q} + \tilde{L}^{\dagger}m_{L}^{2}\tilde{L} + \tilde{\bar{u}}m_{u}^{2}\tilde{\bar{u}}^{\dagger} + \tilde{\bar{d}}m_{d}^{2}\tilde{\bar{d}}^{\dagger} + \tilde{\bar{e}}m_{e}^{2}\tilde{\bar{e}}^{\dagger}\right) \\
& -m_{H_{u}}^{2}H_{u}^{\dagger}H_{u} -m_{H_{d}}^{2}H_{d}^{\dagger}H_{d} + (b H_{d}H_{u} + {\rm c.c.}) \\
& -(A_{u}\bar{Q}H_{u}u + A_{d}\bar{Q}H_{d}d + A_{e}\bar{L}H_{d}e)  \\
& -\mu^\prime {\tilde H_u}\cdot
{\tilde H_d} -\tilde{Q}~{H}_d^{\dagger} A^\prime_{u} \tilde{U}-
\tilde{Q}~ {H}_u^{\dagger} A^\prime_{d} \tilde{D}
 - \tilde{L}~{H}_u^{\dagger} A^\prime_{e} \tilde{E} - \mbox{h.c.}~,
\end{array}
\label{eq:SSBLag}
\end{equation}
where the first line displays the gaugino masses which are induced at one-loop at $M_{{\rm Mess}}$ as

\begin{equation}
M_{i} = \dfrac{\alpha_{i}}{4\pi}\Lambda~,
\label{eq:mgauginos}
\end{equation}
where $i=1,2,3$ corresponds to $U(1)_{Y}$, $SU(2)_{L}$ and $SU(3)_{C}$, respectively. The second and third lines involve the SSB masses for the sfermions and MSSM Higgs fields, respectively, and these masses are induced at two-loop level as:

\begin{equation}
m^2_\phi(\mmess)=2 N_5 \Lambda^2 \sum_{i=1}^{3} C_i(\phi) \left( \frac{\alpha_i}{4\pi}\right)^2,
\label{mscalars}
\end{equation}
where $\phi$ stands for the MSSM fields listed in the second and third lines of Eq.(\ref{eq:SSBLag}). The quadratic Casimir invariant are given by $C_{1} = (3/5)(Y/2)^{2}$, $C_{2} = 3/4$ and $C_{3} = 4/3$, corresponding to the gauge groups of SM. Note that $C_{1}$ depends on the hypercharge $Y$ of these fields.

The fourth line in the SSB Lagrangian is the usual SSB trilinear scalar interactions terms, and as mentioned before, they are generated at the third-loop level. Therefore, their magnitude is suppressed by $M_{{\rm Mess}}$ compared to the scalar mass-squares. Thus, they can be neglected at $M_{{\rm Mess}}$. Note that even if one sets them to zero at $M_{{\rm Mess}}$, they can still have considerable magnitudes at the low scale through their RGE evolution, since the SSB mass terms are effective in their RGEs. 

{Despite being strictly forbidden in a supersymmetric model, the NH terms can be induced by higher dimensional operators through the SUSY breaking as }

\begin{equation}
\setstretch{2.5}
\begin{array}{lll}
\dfrac{1}{M^{3}}\bigint d^{4}x d^{4}\theta \hat{X}\hat{\bar{X}}^{\dagger}\hat{\Phi}_{i}\hat{\Phi}_{j}\hat{\Phi}_{k}^{\dagger} &  = \dfrac{\mathcal{F}^{2}}{M^{3}}S_{i}S_{j}S_{k}^{*} & \rightarrow A^{\prime}S_{i}S_{j}S_{k}^{*}~, \\
\dfrac{1}{M^{3}}\bigint d^{4}x d^{4}\theta \hat{X}\hat{\bar{X}}^{\dagger}\mathcal{D}^{\alpha}\hat{\Phi}\mathcal{D}_{\alpha}\hat{\Phi} & = \dfrac{\mathcal{F}^{2}}{M^{3}}\psi\psi & \rightarrow \mu^{\prime}\psi\psi~,
\end{array}
\label{eq:NHd}
\end{equation}
{which can be adapted to GMSB models by setting $M\rightarrow M_{{\rm Mess}}$ and $\mathcal{F} = \Lambda M_{{\rm Mess}}$. The typical order of magnitude for the NH terms is then found to be}

\begin{equation}
A^{\prime},\mu^{\prime} \sim \dfrac{\Lambda^{2}}{M_{{\rm Mess}}}~,
\label{eq:NHmagd}
\end{equation}
{which implies a relation between the scales of the SUSY scalar masses and the NH terms as $m_{0}^{2} \sim A_{0}^{\prime}M_{{\rm Mess}}$. Therefore, the NH terms are induced at relatively small magnitudes compared to the SUSY scalar mass-squared terms, although Eq.(\ref{eq:NHmagd}) allows the NH terms to be as large as about 200 TeV \cite{Nis:2025fxc}. }

{While the operators in Eq.~(\ref{eq:NHd}) establish a direct link between the NH terms and SUSY breaking, NH mass terms for SUSY fermions (i.e. Higgsinos and gauginos) can be induced through independent mechanisms. For instance, such terms can emerge in $\mathcal{N}=2,4$ SUSY gauge models when the gauge invariant action is decomposed into $\mathcal{N}=1$ fields \cite{deWit:1996kc,Bellisai:1997ck,Henningson:1995eh,Gonzalez-Rey:1998xrq,Buchbinder:1998qd,Antoniadis:2008es}. Alternatively, strongly coupled SUSY gauge dynamics can also be addressed for such NH terms \cite{Arkani-Hamed:1998dti} for the masses of the supersymmetric fermions. Given these possible frameworks, we do not restrict $\mu^{\prime}$ values to those implied by the operators given in Eq.(\ref{eq:NHd}). These NH terms are represented in the last line of Eq.(\ref{eq:SSBLag}), where the indices in $A_{i}^{\prime}$ terms are suppressed, but the third family sfermions are referred to $i=t,b,\tau$.
}

\subsection{NH Impact on YU}
\label{subsec:NHimpYU}

Even though it is imposed at $\mgut$ the YU condition is also quite effective in shaping the low scale implications. This is because YU cannot be compatible even with the third family fermion masses, unless the threshold contributions from the SUSY particles are considered at $M_{{\rm SUSY}}$ at which the SUSY particles are assumed to decouple from the physical spectrum. The earlier studies on YU have revealed that especially $y_{b}$ should receive large negative contributions to satisfy the YU condition consistently \cite{Gogoladze:2010fu}.

\begin{figure}[h!]
\centering
\includegraphics[scale=0.7]{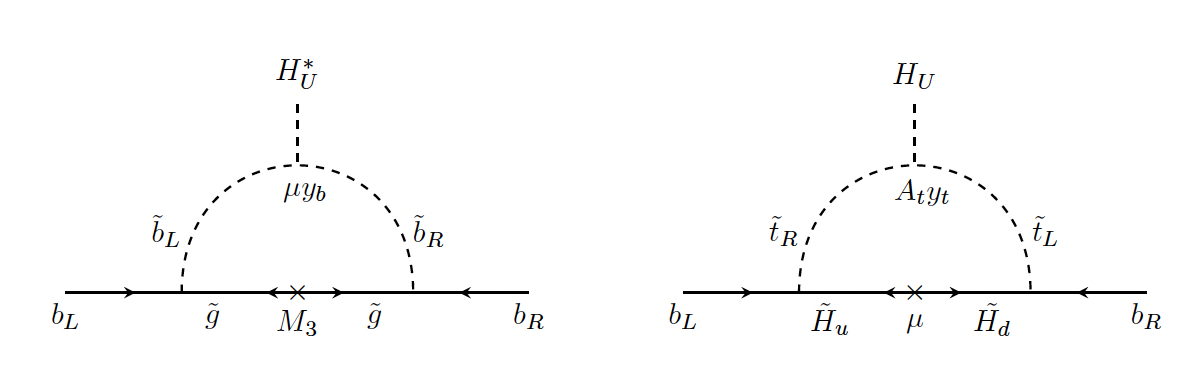}
\caption{The threshold corrections to the b-quark Yukawa coupling.}
\label{fig:ybth}
\end{figure}

The leading order threshold contributions are shown diagrammatically in Figure \ref{fig:ybth}. Note that these diagrams represent only the leading order holomorphic contributions, which can be written as \cite{Hall:1993gn}
\begin{equation}
\delta_{y_{b}}^{H} \approx \dfrac{g_{3}^{2}}{12\pi^{2}}\dfrac{\mu M_{3}\tan\beta}{m_{\tilde{b}}^{2}}+\dfrac{y_{t}^{2}}{32\pi^{2}}\dfrac{\mu A_{t}\tan\beta}{m_{\tilde{t}}^{2}}~~,
\label{eq:deltab}
\end{equation}
where $m_{\tilde{b}}^{2} = m_{\tilde{b}_{L}}m_{\tilde{b}_{R}}$ and $m_{\tilde{t}}^{2} = m_{\tilde{t}_{L}}m_{\tilde{t}_{R}}$. Models with $\mu > 0$ need quite large negative $A_{t}$ terms to provide large negative threshold contributions to $y_{b}$, which does not seem possible to realize in mGMSB. This is one of the reasons that YU prefers the models with $\mu < 0$. Even in those models, heavy mass spectra for strongly interacting particles can significantly suppress these contributions. In addition, models with $\mu < 0$ can be problematic with the current measurement of the muon anomalous magnetic moment (muon $g-2$) \cite{Aliberti:2025beg,Muong-2:2025xyk}, since the dominant SUSY contributions happen to be negative.

The NH contributions to $y_{b}$ can be encoded with the terms in the last line of Eq.(\ref{eq:SSBLag}). These terms contribute through the first diagram in Figure \ref{fig:ybth} proportional to $A_{b}^{\prime}y_{b}$. With such contributions from $A_{b}^{\prime}$, the overall contribution from the first diagrams can turn out to be negative even in the cases with $\mu > 0$. Another dominant contribution arises through the second diagram by enhancing the Higgsino mass. The overall NH threshold contribution to $y_{b}$ can be written as

\begin{equation}
\delta_{y_{b}}^{{\rm NH}} \approx \dfrac{g_{3}^{2}}{12\pi^{2}}\dfrac{A_{b}^{\prime} M_{3}\tan\beta}{m_{\tilde{b}}^{2}}+\dfrac{y_{t}^{2}}{32\pi^{2}}\dfrac{\mu^{\prime} A_{t}\tan\beta}{m_{\tilde{t}}^{2}}~~,
\label{eq:deltabNH}
\end{equation}
and the total threshold contribution becomes $\delta y_{b} = \delta_{y_{b}}^{H} + \delta_{y_{b}}^{NH}$. With these NH contributions, the impact from YU on the holomorphic contributions can be entirely transferred to the interplay among the NH terms; thus one can accommodate the YU solutions in models in which they are  not possible in the absence of NH terms.

{Eq.(\ref{eq:deltabNH}) clearly shows that negative values of $A_{b}^{\prime}$ are favored by the YU condition, and a negative $\mu^{\prime}$ is expected to help minorly. However, the enhancement in the negative threshold corrections to $y_{b}$ from negative values of the NH terms is not straightforward, since these terms can significantly alter the low scale implications. Indeed, their interference is highly subtle, and they can be revealed by studying their effects through the RGEs. For instance, assuming the heavy mass scales for the SUSY scalars prevent them from developing non-trivial VEVs \cite{Demir:2014jqa}, the NH terms do not participate in the electroweak symmetry breaking (EWSB); thus, the EWSB condition remains intact as follows:} 

\begin{equation}
\dfrac{M_{Z}^{2}}{2} = -\mu^{2} - \dfrac{m_{H_{u}}^{2}\tan^{2}\beta - m_{H_{d}}^{2}}{\tan^{2}\beta - 1}~,
\label{eq:ewsbd}
\end{equation}
{where the $m_{H_{u}}^{2}$ and $m_{H_{d}}^{2}$ are the SSB mass-squared terms for the MSSM Higgs doublets. Note that, in our notation, $m_{H_{u}}^{2}$ and $m_{H_{d}}^{2}$ include also the loop contributions. The condition imposed by Eq.(\ref{eq:ewsbd}) leads to $-\mu^{2} \simeq m_{H_{u}}^{2} $ and they cancel each other such that Eq.(\ref{eq:ewsbd}) yields a consistent $Z-$boson mass. The term with $m_{H_{d}}^{2}$ is suppressed by $\tan\beta$, and therefore its contribution is	 negligible. In this context the required cancellation between $\mu^{2}$ and $m_{H_{u}}^{2}$ leads to $m_{H_{u}}^{2} < 0 $ in solutions consistent with the EWSB condition. This fact also restricts the NH terms implicitly. Even though Eq.(\ref{eq:ewsbd}) does not depend on the NH terms explicitly, they can take part in EWSB through the renormalization group equations (RGEs) as 
}
\begin{equation}
\dfrac{d m_{H_{u}}^{2}}{dt} =\left(\dfrac{d m_{H_{u}}^{2}}{dt}\right)_{{\rm MSSM}} +6\left(\dfrac{1}{5}g_{1}^{2} + g_{2}^{2}\right)\mu^{\prime 2} - 6y_{b}^{2}A_{b}^{\prime 2} - 2y_{\tau}^{2}A_{\tau}^{\prime 2}~,
\label{eq:mHuRGE}
\end{equation}
{where $t=\log(Q)$, and the RGE is written for the run from $M_{{\rm Mess}}$ to the electroweak scale. The usual MSSM RGE for $m_{H_{u}}^{2}$ is denoted by the first term, while the rest of the terms reveal the NH impact on $m_{H_{u}}^{2}$ through its RGE. As imposed by the GMSB mass relations given in Eqs.(\ref{mscalars}), $m_{H_{u}}^{2}$ receives positive values at $M_{{\rm Mess}}$, and it has to turn out to be negative at $M_{Z}$ through Eq.(\ref{eq:mHuRGE}) as required by the EWSB condition. In this context, despite being varied arbitrarily, $\mu^{\prime}$ drives $m_{H_{u}}^{2}$ towards the positive values, and the last two terms with $A_{b}^{\prime}$ and $A_{\tau}^{\prime}$ should dominate over $\mu^{\prime}$ to drive $m_{H_{u}}^{2}$ to negative values. However, since $A_{b}^{\prime} = A_{\tau}^{\prime} = A_{0}^{\prime} = \pm\Lambda^{2}/M_{{\rm Mess}}$ at $M_{{\rm Mess}}$, $\mu^{\prime}$ needs to be relatively small. Note that $m_{H_{u}}^{2}$ can be as large as about $100$ TeV$^{2}$ at $M_{{\rm Mess}}$, thus the impact from the NH terms plays a crucial role in satisfying the EWSB condition. This interplay among the NH terms yields negative but small $m_{H_{u}}^{2}$ at $M_{Z}$, which leads to a small $\mu-$term, since the EWSB condition implies $\mid-\mu^{2}\mid \simeq \mid m_{H_{u}}^{2} \mid$. The small values for $\mu$ are preferred by the YU solutions, since in our setup, they lead to rather positive threshold contributions.
}

{The EWSB condition restricts the NH terms regardless of their signs, but as mentioned above, the YU condition favors negative NH terms, and sufficiently large negative values for $A_{0}^{\prime}$ might seem to provide the required negative threshold corrections to $y_{b}$ at first glance. However, a negative $A_{0}^{\prime}$ is not enough to give rise to negative contributions, since these terms also drive the masses of the SUSY scalars to heavier scales \cite{Un:2014afa,Un:2023wws,Nis:2025fxc}. In GMSB models, the NH contributions to sbottoms and stops can increase their masses up to about 6-10 TeV. Therefore, they also strengthen the suppression in the threshold corrections due to $m_{\tilde{b}}$ and $m_{\tilde{t}}$. Note that such effects from the NH terms do not appear when they are effectively imposed at the low scale regardless of their possible origins \cite{Chattopadhyay:2017qvh,Ali:2021kxa,Chakraborty:2019wav,Un:2014afa,Rehman:2022ydc,Un:2023wws,Israr:2025cfd,Israr:2024ubp,Rehman:2024tdr,Rehman:2025djc,Rehman:2025qew}.}

\subsection{NH Impact on muon $g-2$}
\label{subsec:muong2NHd}

The recently confirmed agreement on the muon anomalous magnetic moment (hereafter muon $g-2$) between its experimental measurements and the SM predictions \cite{Muong-2:2025xyk,Aliberti:2025beg} can lead to a considerably strong impact on shaping the NH contributions to the low scale observables. The NH terms rather serve to suppress the supersymmetric contributions to muon $g-2$, which might be favored by the current status of muon $g-2$ in the models that typically lead to large supersymmetric contributions to muon $g-2$ \cite{Nis:2025fxc}. Such a suppression can be even more interesting due to the fact that the solutions are expected to be compatible with the YU condition in regions with negative $A^{\prime}$ terms. In these solutions, the muon $g-2$ contributions can easily turn out to be negative in the minimal GMSB models, since the supersymmetric contributions to muon $g-2$ are nearly zero already without the NH terms due to the heavy mass spectra \cite{Ajaib:2012vc,Gogoladze:2015tfa}.

\begin{figure}[h!]
\centering
\includegraphics[scale=1.5]{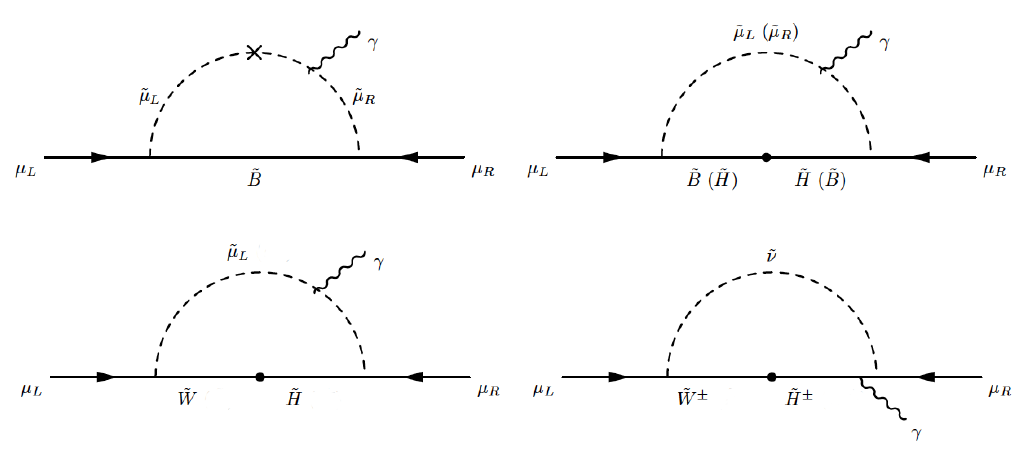}
\caption{The diagrams of leading supersymmetric contributions to muon $g-2$. The cross in the top-left diagram represents the chirality flip between the left- and right-handed smuons, while the dots in the other diagrams denote the mixing among the MSSM neutralino species.}
\label{fig:muong2diags}
\end{figure}

The leading supersymmetric contributions to muon $g-2$ can be classified in four classes which are diagrammatically shown in Figure \ref{fig:muong2diags}. The cross in the top-left diagram represents the chirality flip between the left- and right-handed smuons, while the dots in the other diagrams denote the mixing among the MSSM neutralino species. The contributions from each diagram (ordered from top-left to bottom right, respectively) can be calculated in the presence of the NH terms as follows \cite{Moroi:1995yh,Martin:2001st,Giudice:2012pf,Fargnoli:2013zia}:

\begin{equation}\hspace{-2.3cm}
\Delta a_{\mu}^{\tilde{B}\tilde{\mu}_{L}\tilde{\mu}_{R}} \simeq \dfrac{g_{1}^{2}}{16\pi^{2}}\dfrac{m_{\mu}^{2}M_{\tilde{B}}\left[(\mu + A_{\mu}^{\prime})\tan\beta - A_{\mu}\right]}{m_{\tilde{\mu}_{L}}^{2}m_{\tilde{\mu}_{R}}^{2}}F_{N}\left( \dfrac{m_{\tilde{\mu}_{L}}^{2}}{M_{\tilde{B}}^{2}},\dfrac{m_{\tilde{\mu}_{R}}^{2}}{M_{\tilde{B}}^{2}}\right)~, \tag{2.16-a}
\label{eq:binocont}
\end{equation}

\begin{equation}
\setstretch{3.5}
\begin{array}{ll}
\Delta a_{\mu}^{\tilde{B}\tilde{H}\tilde{\mu}} \simeq  \dfrac{g_{1}^{2}}{16\pi^{2}}m_{\mu}^{2}M_{\tilde{B}}\left[(\mu + A_{\mu}^{\prime})\tan\beta - A_{\mu}\right] & \left[
\dfrac{1}{m_{\tilde{\mu}_{L}}^{4}}F_{N}\left(\dfrac{M_{1}^{2}}{m_{\tilde{\mu}_{L}}^{2}}, \dfrac{\mu^{2}}{m_{\tilde{\mu}_{L}}^{2}}   \right) \right.\\
& - \left. \dfrac{2}{m_{\tilde{\mu}_{R}}^{4}}F_{N}\left(\dfrac{M_{1}^{2}}{m_{\tilde{\mu}_{R}}^{2}}, \dfrac{\mu^{2}}{m_{\tilde{\mu}_{R}}^{2}}   \right) \right]~,
\end{array} \tag{2.16-b}
\label{eq:bhcont}
\end{equation}

\vspace{0.3cm}
\begin{equation}\hspace{-2.7cm}
\Delta a_{\mu}^{\tilde{W}\tilde{H}\tilde{\mu}} \simeq  \dfrac{g_{2}^{2}}{16\pi^{2}} \dfrac{m_{\mu}^{2}M_{2}\left[(\mu + A_{\mu}^{\prime})\tan\beta - A_{\mu}\right]}{m_{\tilde{\mu}_{L}}^{4}}F_{N}\left(\dfrac{M_{2}^{2}}{m_{\tilde{\mu}_{L}}^{2}}, \dfrac{\mu^{2}}{m_{\tilde{\mu}_{L}}^{2}}   \right)~, \tag{2.16-c}
\label{eq:whcont}
\end{equation}

\vspace{0.3cm}
\begin{equation}\hspace{-3.2cm}
\Delta a_{\mu}^{\tilde{W}\tilde{H}\tilde{\nu}} \simeq \dfrac{g_{2}^{2}}{8\pi^{2}}\dfrac{m_{\mu}^{2}M_{2}\left[(\mu + A_{\mu}^{\prime})\tan\beta - A_{\mu}\right]}{m_{\tilde{\nu}}^{4}}F_{C}\left(\dfrac{M_{2}^{2}}{m_{\tilde{\nu}}^{2}},\dfrac{\mu^{2}}{m_{\tilde{\nu}}^{2}} \right)~, \tag{2.16-d}
\label{eq:charcount}
\end{equation}
where 

\begin{equation}
F_{N}(x,y) = xy\left[\dfrac{-3+x +y + xy}{(x-1)^{2}(y-1)^{2}} + \dfrac{2x\log(x)}{(x-y)(x-1)^{3}} -\dfrac{2y\log(y)}{(x-y)(y-1)^{3}}\right]~,\tag{2.17-a}
\label{eq:loopN1}
\end{equation}

\begin{equation}
F_{C}(x,y)=xy\left[ \dfrac{5-3(x+y)+xy}{(x-1)^{2}(y-1)^{2}} - \dfrac{2\log(x)}{(x-y)(x-1)^{3}}+\dfrac{2\log(y)}{(x-y)(y-1)^{3}}\right]~.\tag{2.17-b}
\label{eq:loopN2}
\end{equation}
\setcounter{equation}{17}
Note that the loop functions are explicitly defined  such that their arguments reflect the heavy mass limits for the corresponding particles.

The dominant supersymmetric contributions to muon $g-2$, in general, arise from the top-left diagram approximated by Eq.(\ref{eq:binocont}), since they are induced through gauge interactions and enhanced by the chirality flip between the smuons. The contributions from the other diagrams involve the Yukawa coupling between smuons and Higgsinos ($\simeq 10^{-3}$), thus they are expected to be minor compared to the bino-smuon loop. In the presence of the NH terms, the sign of the dominant contributions is determined by the factor $(\mu + A_{\mu}^{\prime})\tan\beta$ which restricts negative $A_{\mu}^{\prime}$ values to $\lvert A_{\mu}^{\prime}\rvert \lesssim \mu$ such that the supersymmetric muon $g-2$ contributions remain small and positive. Indeed, this restriction is necessary for all the contributions to keep them positive except for those arising from the right-handed smuon-Bino-Higgsino loop, whose magnitude is determined also by the second loop factor in the parentheses of Eq.(\ref{eq:bhcont}). These contributions turn out to be positive when $\lvert A_{\mu}^{\prime}\rvert \gtrsim \mu$ in the regions with negative $A_{\mu}^{\prime}$. Furthermore, they can dominate over the other contributions when the right-handed smuon is sufficiently lighter than the left-handed smuon.

In this context, if the top-left diagram is identified as the dominant contributions, the solutions consistent with the current measurements of muon $g-2$ impose a restrictive impact on the YU solutions. On the other hand, it might still be possible to realize such solutions with large and negative $A_{\mu}^{\prime}$ terms, if the right-handed smuon-bino-higgsino loop dominates over the other diagrams. In this case, the contributions lead to different scenarios, since they involve interactions between smuons and higgsinos, whose strength is determined by the Yukawa coupling associated with the second family of leptons ($y_{\mu}\sim 10^{-3}$). At first glance, this requires relatively light Higgsinos, which makes the NH $\mu^{\prime}$ term also relevant in the muon $g-2$ implications. Before concluding, we should also note that negative $A_{\mu}^{\prime}$ terms cannot still be arbitrarily large, since they drive the smuon masses to heavy scales \cite{Nis:2025fxc}.

\section{Scanning Procedure and Experimental Constraints}
\label{sec:scan}
In this section, we briefly describe the fundamental parameter space of the models in the mGMSB class, and the scanning procedure which is optimized to generate statistically well-distributed YU solutions. The fundamental parameters and their ranges in our scans are the following:

\begin{equation}
\setstretch{1.5}
\begin{array}{rcl}
10^3 \leq & \Lambda & \leq 10^8 ~{\rm GeV}, \\
10^6 \leq & M_{\rm Mess} & \leq 10^{16} ~{\rm GeV}, \\
35 \leq & \tan\beta & \leq 60, \\
-20 \leq & \mu^{\prime} & \leq 20~{\rm TeV}~,\\
& {\rm sgn}(A^{\prime}_{0}) & =-1,+1~,
\end{array}
\label{eq:GMSB_NH}
\end{equation}
where $\Lambda$ and $M_{{\rm Mess}}$ determine the SSB masses as discussed in the previous section, and $\tan\beta$ is the usual MSSM parameter which is defined in terms of the VEVs of the MSSM Higgs fields as $\tan\beta \equiv v_{u}/v_{d}$. {Since $\mu^{\prime}$ contributes to the Higgsino mass together with $\mu$, we set its range to the scales comparable with $\mu$ as specified in Eq.(\ref{eq:GMSB_NH}).} The magnitude of $A^{\prime}_{0}$ is determined by $\Lambda$ and $M_{{\rm Mess}}$ as given in Eq.(\ref{eq:NHmagd}), but we also allow that it can have relative sign due to phase differences, and we randomly assign its sign in our scans.

We perform random scans over the fundamental parameters listed above by employing the Metropolis-Hastings algorithm \cite{Baer:2008jn,Belanger:2009ti}. Their randomly determined values are inputted into SPheno-4.0.4 \cite{Porod:2003um,Goodsell:2014bna}, which calculates the mass spectra, mixing of the supersymmetric and Higgs fields, branching ratios of their decays etc. This numerical calculation package is generated by SARAH \cite{Staub:2008uz,Staub:2015iza}, which is also modified such that the NH terms are imposed at $M_{{\rm Mess}}$, and they are involved with the other SSB terms in the RGEs. SPheno, by default, involves the RGEs below $M_{{\rm Mess}}$, but we supply it with the RGEs between $M_{{\rm Mess}}$ and $\mgut$, which are given in Eqs.(\ref{eq:abRGE} and \ref{eq:yuRGE}). $\mgut$ is determined dynamically by the condition $g_{1}=g_{2}\simeq g_{3}$, where $g_{1}$, $g_{2}$ and $g_{3}$ denote the gauge couplings of $U(1)_{Y}$, $SU(2)_{L}$ and $SU(3)_{c}$, respectively. In the unification condition, we do not require a strict equality by allowing small deviation in $g_{3}$ to take into account some unknown GUT scale threshold corrections \cite{Hisano:1992jj,Yamada:1992kv,Chkareuli:1998wi}. In the RGE evolutions, the decoupling scale of the SUSY particles is determined by $M_{{\rm SUSY}} = \sqrt{m_{\tilde{t}_{L}}m_{\tilde{t}_{R}}}$, and the threshold contributions to the Yukawa couplings discussed in Section \ref{subsec:NHimpYU} are calculated and added at this scale. 

In our analyses, we subsequently apply several theoretical and experimental constraints. One of the main constraints arises from the radiative electroweak symmetry breaking (REWSB) condition which restricts the values of $\mu$ and $b$ with the SSB masses of the Higgs fields $m_{H_{u}}$ and $m_{H_{d}}$. In our scans, we accept only the solutions compatible with the REWSB condition. After generating the data, we first apply the mass bounds on the SM-like Higgs boson \cite{ATLAS:2012yve,CMS:2012qbp,CMS:2013btf} and SUSY particles \cite{ParticleDataGroup:2014cgo,ATLAS:2021twp,ATLAS:2020syg,ATLAS:2022rcw}, the constraints from rare decays of $B-$meson \cite{Belle-II:2022hys,CMS:2020rox}. We also constrain the muon $g-2$ results, since the current experimental \cite{Muong-2:2025xyk} and theoretical \cite{Aliberti:2025beg} analyses have revealed a significant agreement for the SM predictions. The constraints leading to main impacts in our analyses can be listed as 

\begin{equation}
\setstretch{1.8}
\begin{array}{l}
m_h  = 123-127~{\rm GeV}~,\\
m_{\tilde{g}} \geq 2.1~{\rm TeV}~(1400~{\rm GeV}~{\rm if~it~is~NLSP})~,\\
1.95\times 10^{-9} \leq{\rm BR}(B_s \rightarrow \mu^+ \mu^-) \leq 3.43 \times10^{-9} \;(2\sigma)~, \\
2.99 \times 10^{-4} \leq  {\rm BR}(B \rightarrow X_{s} \gamma)  \leq 3.87 \times 10^{-4} \; (2\sigma)~, \\
0 \leq  \Delta a_{\mu}  \leq 6.6 \times 10^{-10}~.
\label{eq:constraints}
\end{array}
\end{equation}

{While Eq.(\ref{eq:constraints}) lists the constraints which yield a considerable impact on the parameter space, it does not represent the full list of constraints employed in our analyses. We also employ the model independent LEP constraints on the mass spectrum \cite{ALEPH:2002gap,LEPWorkingGroupforHiggsbosonsearches:2003ing,ALEPH:2013dgf} as well as those on the decay modes of the SM Higgs boson including its invisible decay channels \cite{CMS:2023sdw}. In addition, we also require the solutions to be compatible with the electroweak precision observables by bounding the oblique parameters ($S,T,U$) \cite{Haber:2010bw} with the global fits to $W$ and $Z$ boson data \cite{ParticleDataGroup:2024cfk}.} One of the main issues in the theoretical calculations arises in the SM-like Higgs boson predictions. Despite its impressively precise experimental measurement \cite{ParticleDataGroup:2024cfk}, we allow about 2 GeV theoretical uncertainty in applying the Higgs boson mass constraint. This uncertainty of about 2 GeV arises from the top quark mass and strong gauge coupling measurements, as well as the mixing in the squark sector \cite{Allanach:2004rh,Bahl:2019hmm,Bagnaschi:2017xid,Athron:2016fuq,Drechsel:2016htw,Bahl:2020tuq}.

The constraints from rare decays of $B-$meson does not only suppress the large SUSY contributions to the relevant processes, but they also put a strict control in the Higgs sector. The heavy CP-even ($H$) and CP-odd ($A$) higgs bosons mediate the $B_{s}\rightarrow \mu^{+}\mu^{-}$ process \cite{Choudhury:1998ze,Babu:1999hn}. The allowed ranges for these processes are obtained from highly sensitive analyses as given in Eq.(\ref{eq:constraints}). These ranges can drive the Higgs bosons to heavy mass scales, but also they provide a fit on the couplings between the Higgs bosons and SM fermions.  Similarly the $B\rightarrow X_{s}\gamma$ receives contributions from the charged Higgs boson, and the analyses on this process also lead to a fit for the couplings between the Higgs bosons and quarks. It should also be noted that we parametrize the magnitudes of NH trilinear scalar interactions terms as $T_{i}^{\prime} = A_{i}^{\prime}y_{i}$ with $i=u,d,e$ to keep the CKM matrix aligned with the rotations of the squarks.

\section{YU in the Holomorphic Cases}
\label{sec:YUH}
\begin{figure}[h!]
\centering
\begin{subfigure}{0.5\textwidth}
\centering
{\Large $\mu > 0$}
\includegraphics[scale=0.4]{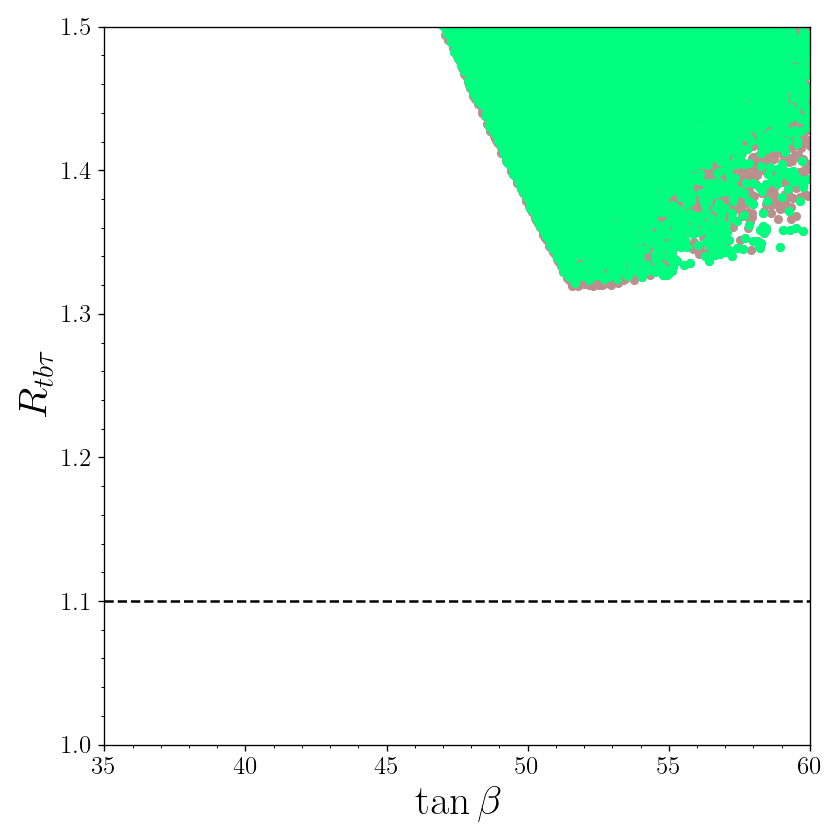}
\end{subfigure}%
\begin{subfigure}{0.5\textwidth}
\centering
{\Large $\mu < 0$}
\includegraphics[scale=0.4]{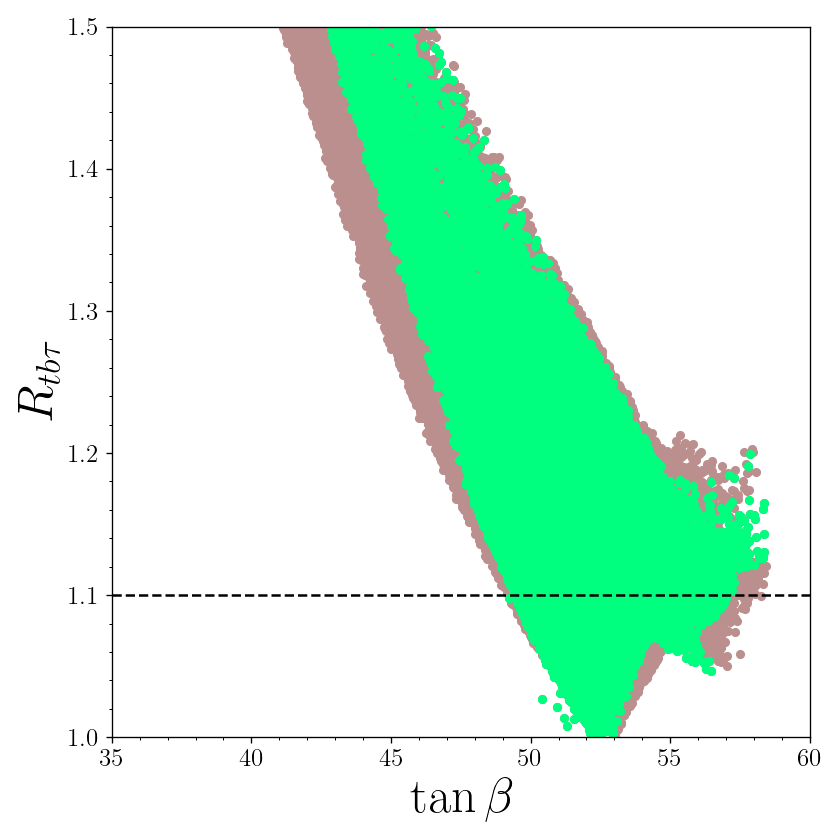}%
\end{subfigure}
\caption{The status of $t-b-\tau$ YU in $R_{tb\tau}-\tan\beta$ planes for the holomorphic cases with $\mu > 0$ (left) and $\mu < 0$ (right). All points are consistent with the REWSB condition. The green points are consistent with the mass bounds and rare decays of $B-$meson given in Eq.(\ref{eq:constraints}). The horizontal dashed line represents the solutions with $R_{tb\tau} = 1.1$, and the points underneath these lines are identified as the YU compatible solutions.}
\label{fig:YUH}
\end{figure}

Before we proceed to discussions of the NH impact on YU in mGMSB, we first display the current status of YU in $R_{tb\tau}-\tan\beta$ planes of Figure \ref{fig:YUH} for the holomorphic cases with $\mu > 0$ (left) and $\mu < 0$ (right). All points are consistent with the REWSB condition. The green points are consistent with the mass bounds and rare decays of $B-$meson given in Eq.(\ref{eq:constraints}). The horizontal dashed line represents the solutions with $R_{tb\tau} = 1.1$, and the points underneath these lines are identified as the YU compatible solutions. As seen from the left plane, the holomorphic cases with $\mu > 0$ cannot accommodate YU within their fundamental parameter space, and $R_{tb\tau}$ can only reach values as low as only about 1.3. This is because, the required negative contributions to $y_{b}$ needs large $A_{t}$ term, which is typically negligible in GMSB models. Therefore, the threshold contributions from the second diagram in Figure \ref{fig:ybth} also become negligible, and $y_{b}$ receives mostly positive contributions from the first diagram when $\mu > 0$. 

\begin{figure}[h!]
\centering
{\Large $\mu < 0 $}
\includegraphics[scale=0.4]{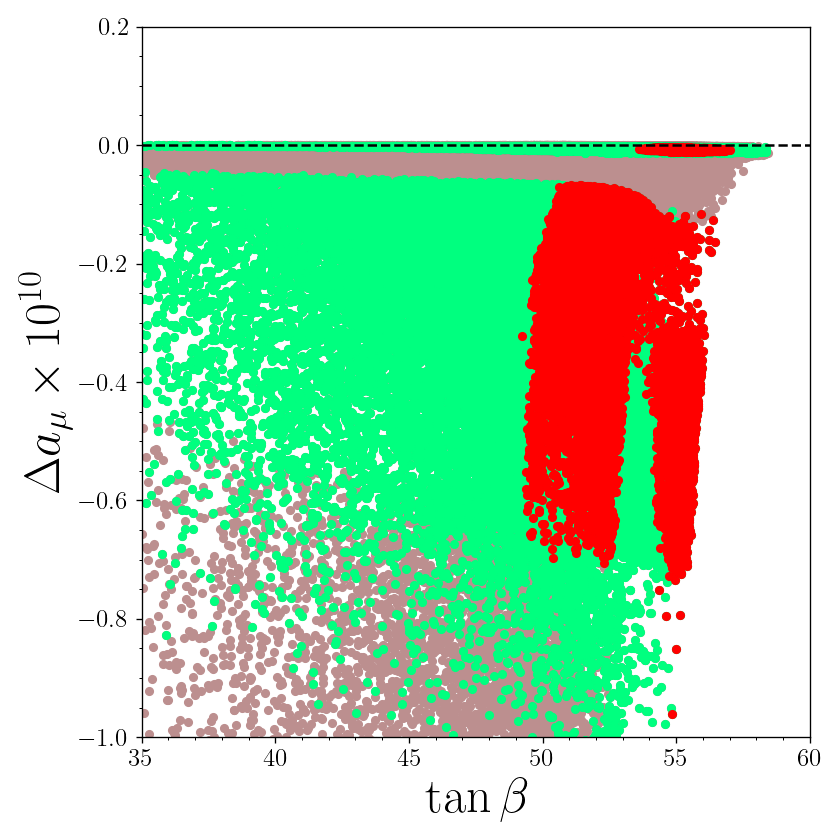}%
\includegraphics[scale=0.4]{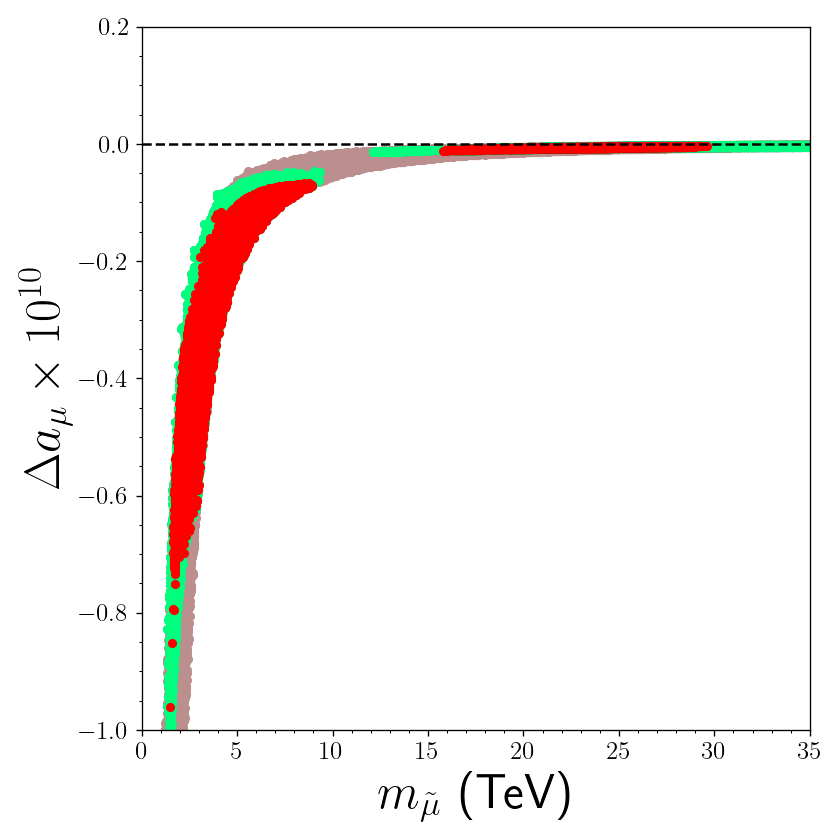}
\includegraphics[scale=0.4]{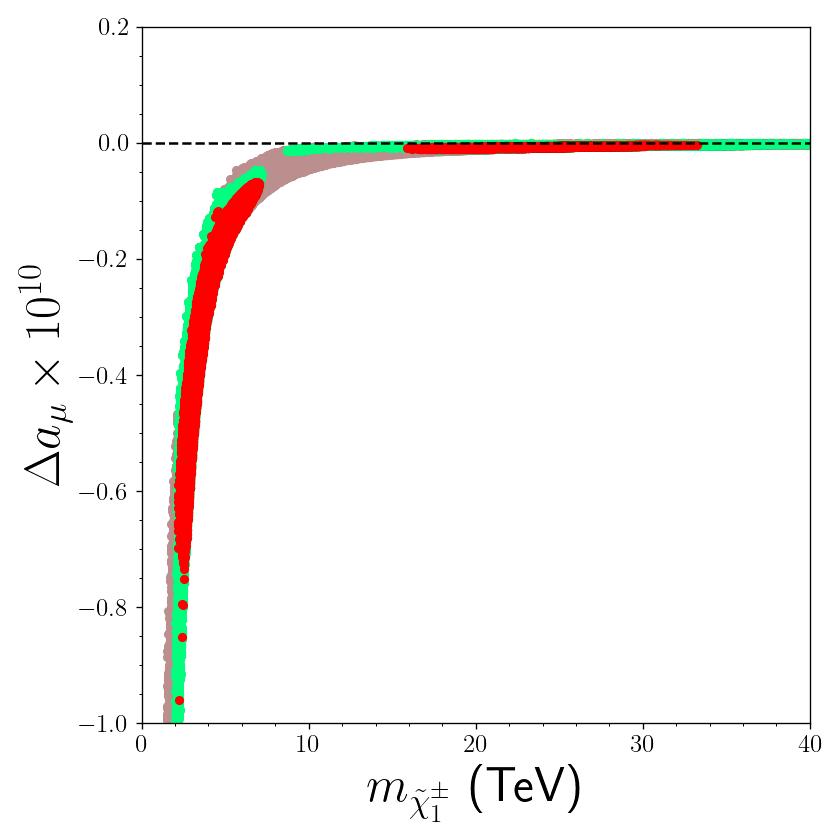}%
\includegraphics[scale=0.4]{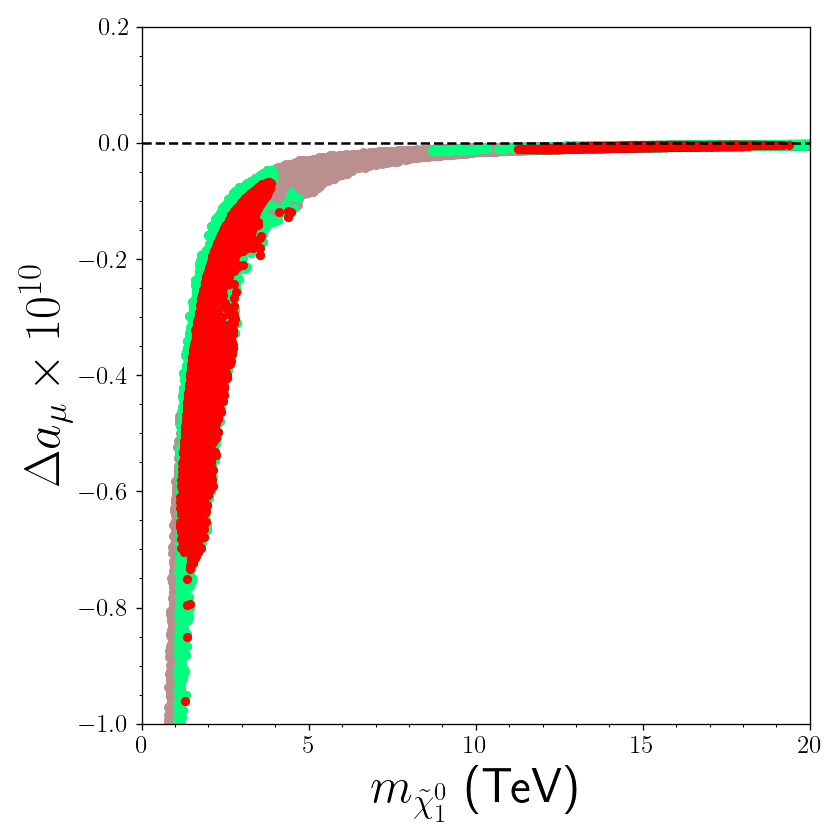}
\caption{Muon $g-2$ ($\Delta a_{\mu}$ results compatible with the YU condition in the case with $\mu < 0$) in the $\Delta a_{\mu}-\tan\beta$, $\Delta a_{\mu} - m_{\tilde{\mu}_{1}}$, $\Delta a_{\mu} - m_{\tilde{\chi}_{1}^{\pm}}$ and $\Delta a_{\mu}-m_{\tilde{\chi}_{1}^{0}}$ planes. All points are compatible with the REWSB condition. The green points represent the solutions consistent with the mass bounds and constraints from rare $B-$meson decays. The red points form a subset of green and they display the YU compatible solutions.}
\label{fig:YUg2}
\end{figure}

The second diagram still remains negligible, but one can realize the desired negative contributions from the first diagram in the holomorphic cases with $\mu < 0$. The right panel in Figure \ref{fig:YUH} shows the results for these cases, and as can be seen from the results, the $\mu < 0$ case can abundantly accommodate the YU solutions. However, these solutions can be problematic when they are confronted with the current results for muon $g-2$. As is well known, the solutions with $\mu < 0$ lead to negative SUSY contributions to muon $g-2$. Even though the heavy spectra of GMSB can significantly suppress such contributions, the current sensitivity of the theoretical calculations and experimental measurements can yield a strong negative impact.

We display our results for muon $g-2$ in the holomorphic cases with $\mu < 0$ in Figure \ref{fig:YUg2} in the $\Delta a_{\mu}-\tan\beta$, $\Delta a_{\mu} - m_{\tilde{\mu}_{1}}$, $\Delta a_{\mu} - m_{\tilde{\chi}_{1}^{\pm}}$ and $\Delta a_{\mu}-m_{\tilde{\chi}_{1}^{0}}$ planes. All points are compatible with the REWSB condition. The green points represent the solutions consistent with the mass bounds and constraints from rare $B-$meson decays. The red points form a subset of green and they display the YU compatible solutions. As is seen all solutions are accumulated in the regions of negative SUSY contributions and the YU solutions can worsen the problem by contributing to muon $g-2$ up to about $-0.8\times 10^{-10}$. Such large negative contributions can be realized in any value of $\tan\beta$ in its YU compatible range (red), while the masses of the smuon, chargino and neutralino play a crucial role in alleviating the muon $g-2$ problem. When these particles are lighter than about 2 TeV, they can significantly worsen the muon $g-2$ results. As seen from the planes of Figure \ref{fig:YUg2}, one needs to constrain these particles to be quite heavier to minimize the SUSY contributions. As seen from the mass planes in Figure \ref{fig:YUg2}, $\Delta a_{\mu}$ can be nearly zero when $m_{\tilde{\mu}}\gtrsim 15$ TeV, $m_{\tilde{\chi}_{1}^{\pm}}\gtrsim 16$ TeV and $m_{\tilde{\chi}_{1}^{0}}\gtrsim 12$ TeV, which are extremely far from the reach of the current and near future collider experiments.

\section{NH Impact on YU in mGMSB with $\mathbf{\mu > 0}$}
\label{sec:NHimpYU}

As discussed in the previous section, the minimal GMSB models can barely be compatible with the YU condition. YU needs to be significantly broken in the $\mu > 0$ case, while it conflicts with the latest muon $g-2$ results in the $\mu < 0$ case. This conflict can be ameliorated when the SUSY mass spectrum is so heavy that it can be beyond the sensitivity even of the future collider experiments. In this section, we discuss the YU solutions compatible with the experimental constraints employed within our analyses in the fundamental parameter space of mGMSB. Note that the green points in the plots of Section \ref{sec:YUH} were not required to satisfy the muon $g-2$ constraint, since we discussed it separately. However, the green points in the rest of our discussions will be required to be consistent with the muon $g-2$ constraint as given in Eq.(\ref{eq:constraints}). After presenting the YU compatible regions in the mGMSB framework, we discuss the YU compatible SUSY mass spectra and prospects to probe YU in the current and/or future collider experiments.


\subsection{The Fundamental Parameter Space of YU}
\label{subsec:FPSYU}

\begin{figure}[h!]
\centering
\includegraphics[scale=0.4]{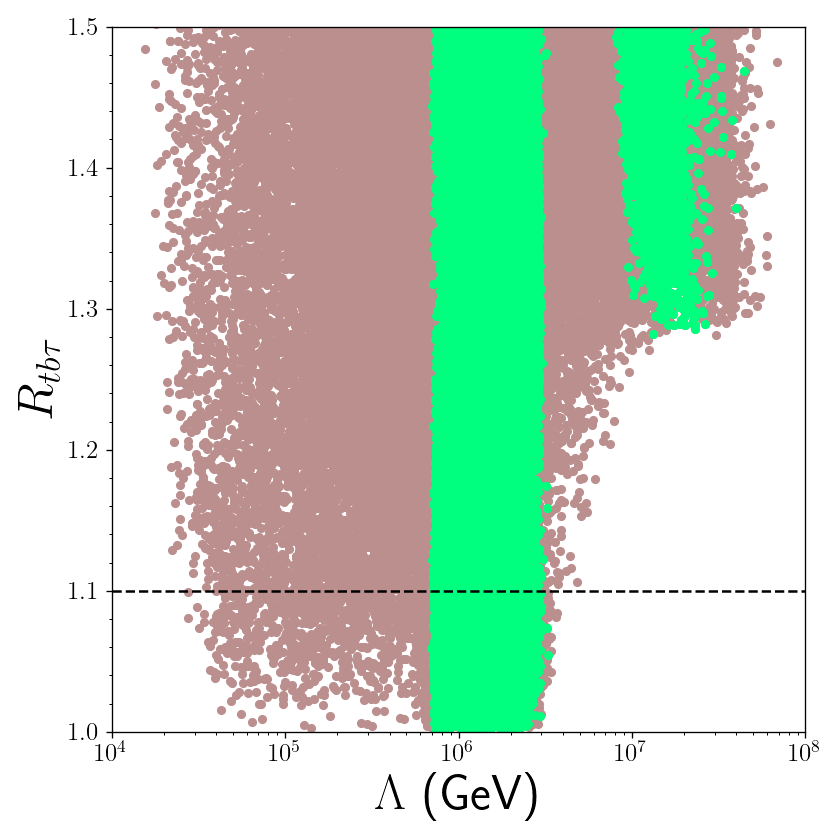}%
\includegraphics[scale=0.4]{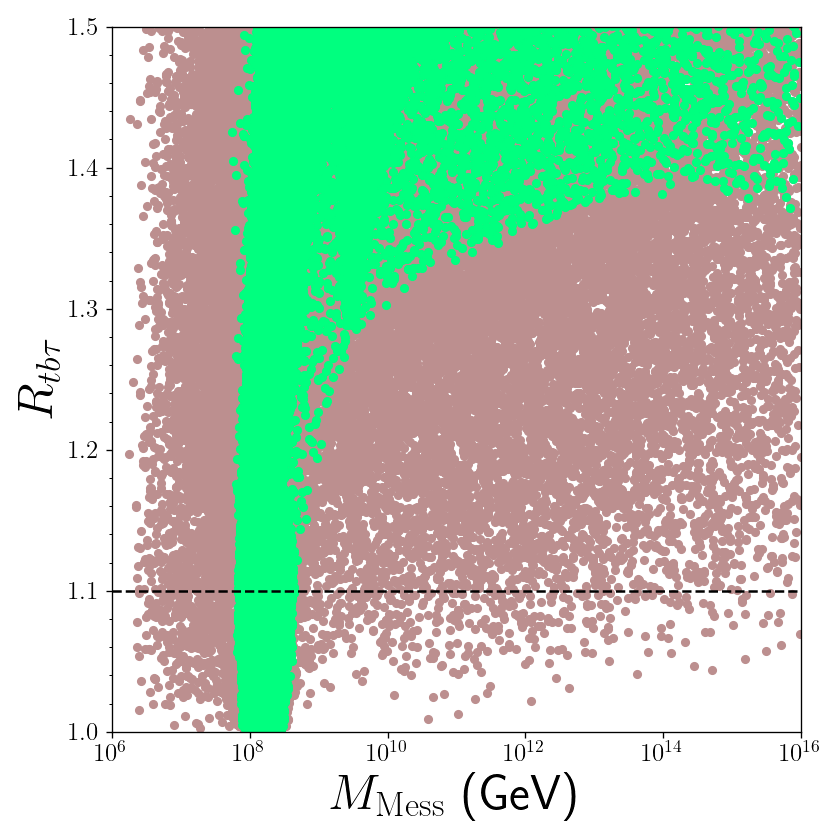}
\includegraphics[scale=0.4]{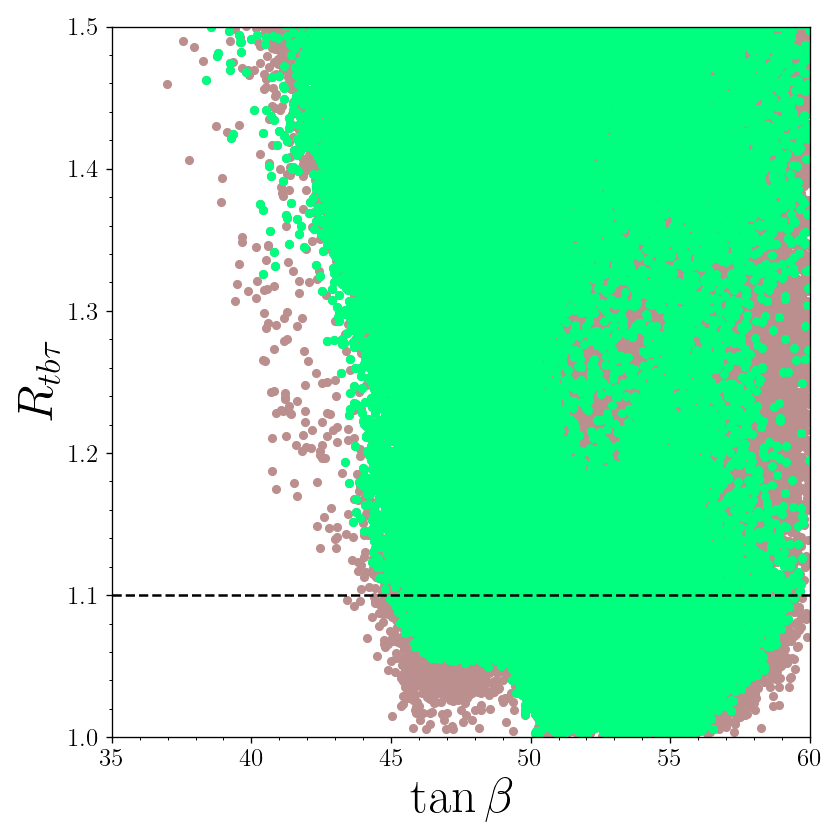}%
\includegraphics[scale=0.4]{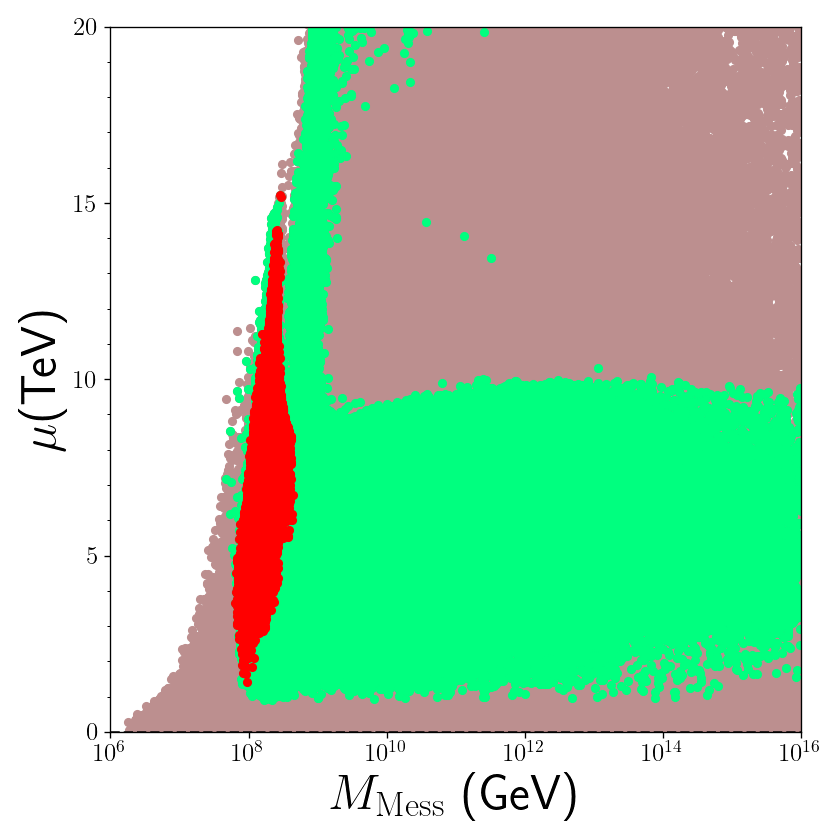}
\caption{The YU solutions in the $R_{tb\tau}-\Lambda$, $R_{tb\tau}-M_{{\rm Mess}}$, $R_{tb\tau}-\tan\beta$ and $\mu - M_{{\rm Mess}}$ planes. The color coding is the same as that described for Figure \ref{fig:YUH}. In the $\mu - M_{{\rm Mess}}$ plane, the red points form a subset of green and they represent the YU solutions, which are compatible with the constraints given in Eq.(\ref{eq:constraints}).}
\label{fig:YUHfund}
\end{figure}

We first display the YU parameter $R_{tb\tau}$ in correlation with various fundamental parameters in the $R_{tb\tau}-\Lambda$, $R_{tb\tau}-M_{{\rm Mess}}$, $R_{tb\tau}-\tan\beta$ and $\mu - M_{{\rm Mess}}$ planes. The color coding is the same as that described for Figure \ref{fig:YUH}. Note that these points are also required to be consistent with the muon $g-2$ constraint. In the $\mu - M_{{\rm Mess}}$ plane, the red points form a subset of green and they represent the YU solutions, which are compatible with the constraints given in Eq.(\ref{eq:constraints}). The top panels show that the fundamental parameters $\Lambda$ and $M_{{\rm Mess}}$ are restricted to narrow ranges as $7\times 10^{5} \lesssim \Lambda \lesssim 2\times 10^{6}$ GeV and $8\times 10^{7} \lesssim M_{{\rm Mess}} \lesssim 8\times 10^{8}$ GeV (green points under the horizontal dashed lines). On the other hand, these regions can be identified with $\tan\beta$ in a range ($45-60$) which is relatively wider than the YU compatible regions in the absence of NH terms as shown in the $R_{tb\tau}-\tan\beta$ plane. We also display the possible ranges of the $\mu-$term compatible with the YU condition (red) in the $\mu - \mmess$ plane. YU, in general, favors large values of $\mu$ ($\gtrsim 3$ TeV) \cite{Gogoladze:2015tfa}, our results can accommodate YU solutions with $\mu \gtrsim 1.8$ TeV.

\begin{figure}[h!]
\centering
\includegraphics[scale=0.4]{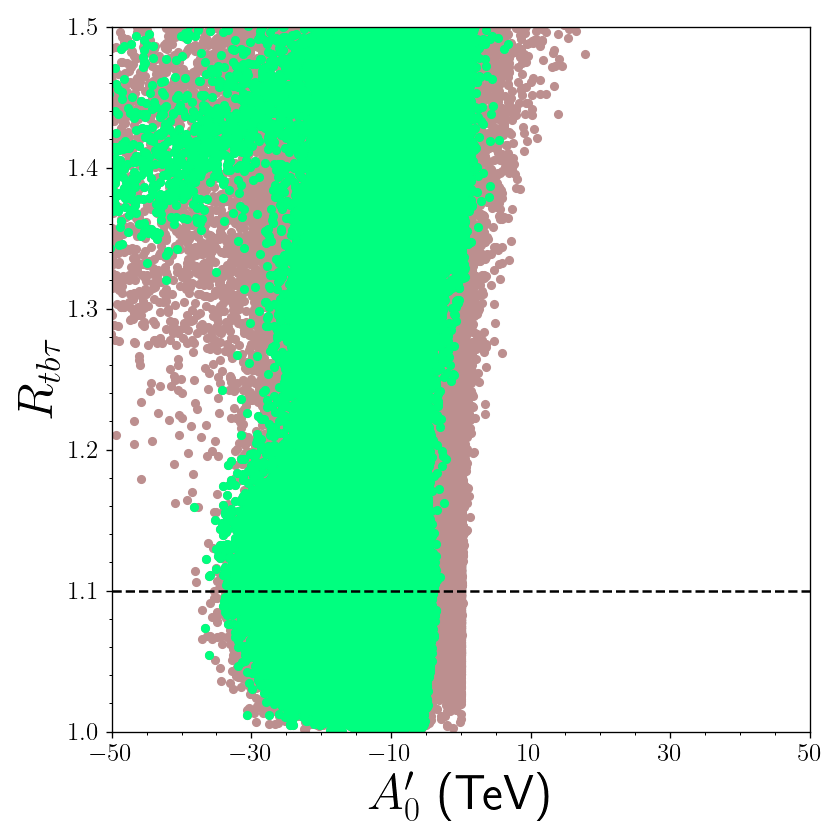}
\includegraphics[scale=0.4]{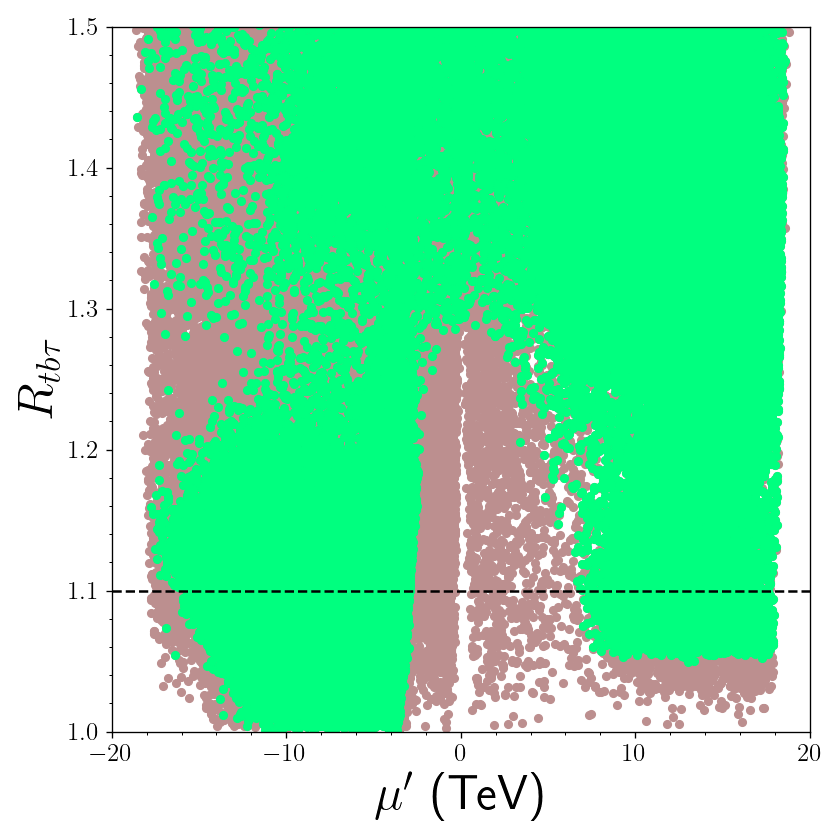}
\caption{The correlation between YU parameter and the NH terms in the $R_{tb\tau}-A_{0}^{\prime}$ and $R_{tb\tau}-\mu^{\prime}$ planes. The color coding and dashed lines are the same as those described for the top panels of Figure \ref{fig:YUHfund}.}
\label{fig:YUNHfund}
\end{figure}

The narrow range for $\Lambda$ and $M_{{\rm Mess}}$ is a direct results of needing to induce considerably effective NH terms as discussed earlier. These terms become almost zero when $M_{{\rm Mess}} \gtrsim 10^{10}$ GeV \cite{Nis:2025fxc}, and the models reduce to the usual GMSB models, which do not involve NH terms. Figure \ref{fig:YUNHfund} displays the ranges for these NH terms in the  $R_{tb\tau}-A_{0}^{\prime}$ and $R_{tb\tau}-\mu^{\prime}$ planes. The color coding and dashed lines are the same as those described for the top planes of Figure \ref{fig:YUHfund}. As seen from the $R_{tb\tau}-A_{0}^{\prime}$ plane, the NH trilinear interaction term is required to be negative and it can take values from about $-40$ to $-8$ TeV. $\mu^{\prime}$ also helps in accommodating the YU solutions, but the impact from YU on this NH term is not as strong as the other parameters. Although perfect YU solutions rather favor the regions with $\mu^{\prime} < 0$, it is still possible to realize the YU solutions when $\mu^{\prime} > 0$ as shown in the $R_{tb\tau}-\mu^{\prime}$ plane.

\begin{figure}[h!]
\centering
\includegraphics[scale=0.4]{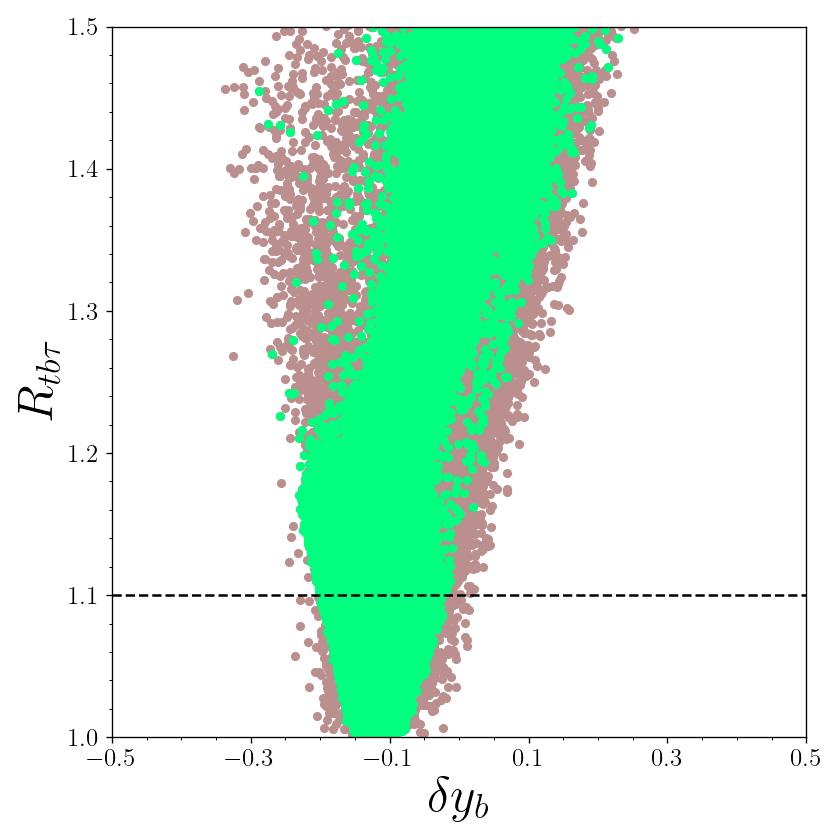}%
\includegraphics[scale=0.4]{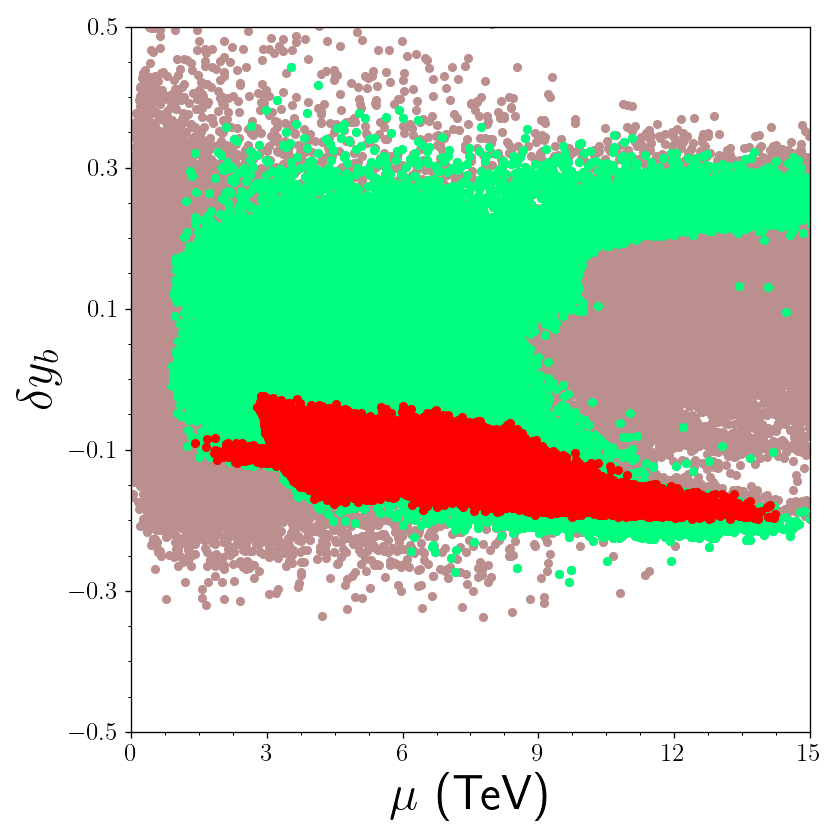}
\includegraphics[scale=0.4]{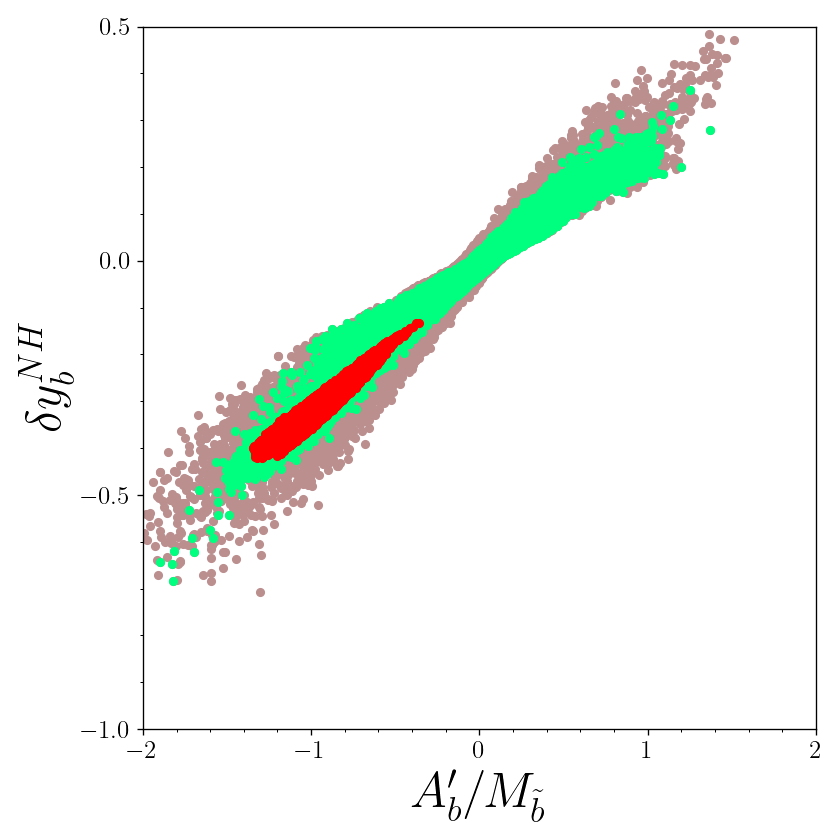}%
\includegraphics[scale=0.4]{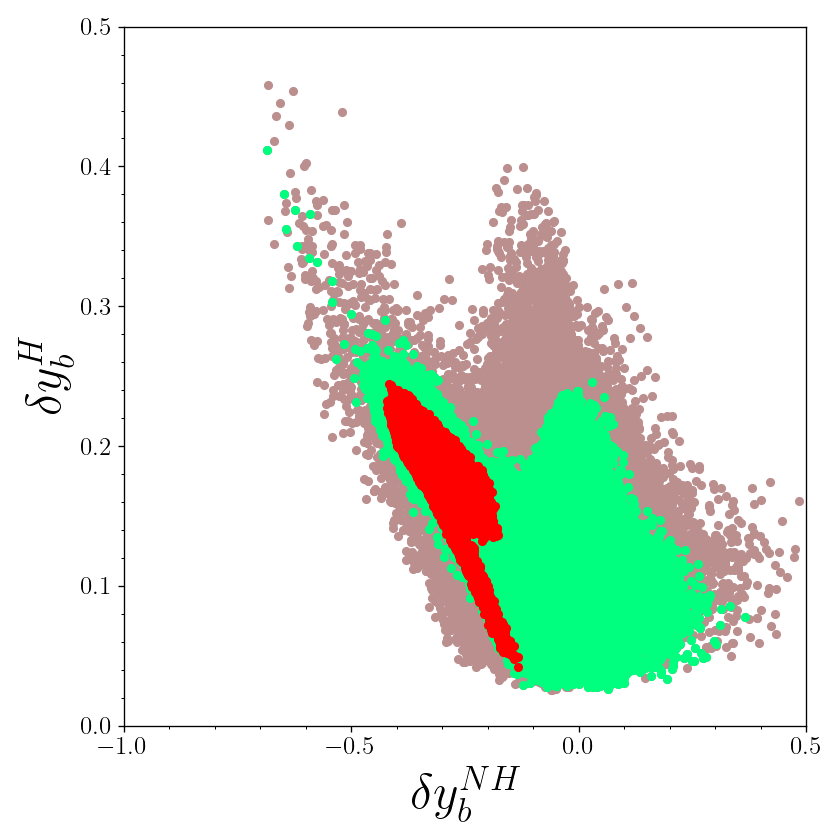}
\caption{The threshold contributions to $y_{b}$ compatible with YU in the $R_{tb\tau}-\delta y_{b}$, $\delta y_{b}-\mu$, $\delta y_{b}-A_{b}^{\prime}/m_{b}$ and $\delta yb^{H}-\delta y_{b}^{NH}$ planes. The color coding and the dashed line in the top-left plane are the same as in Figure \ref{fig:YUH}, while the other planes follow the color coding as given for the bottom-right plane of Figure \ref{fig:YUHfund}.}
\label{fig:Thresh}
\end{figure}

The impact from YU on the NH terms can be understood by considering the threshold contributions to $y_{b}$, which are required to be about $-0.2$ \cite{Gogoladze:2010fu}. As discussed in Section \ref{sec:intro}, the mGMSB models with $\mu > 0$ cannot provide large enough threshold contributions to $y_{b}$ such that YU cannot be realized. On the other hand, when the NH terms are present, they can dominantly adjust the threshold contributions, and the holomorphic terms can even become totally independent on the YU condition. The plots in the $R_{tb\tau}-\delta y_{b}$, $\delta y_{b}-\mu$, $\delta y_{b}-A_{b}^{\prime}/m_{b}$ and $\delta yb^{H}-\delta y_{b}^{NH}$ planes of Figure \ref{fig:Thresh} display the threshold contributions to $y_{b}$. The color coding and the dashed line in the top-left plane are the same as in Figure \ref{fig:YUH}, while the other planes follow the color coding as given for the bottom-right plane of Figure \ref{fig:YUHfund}. As mentioned before, the total threshold corrections should be in the range $-0.25 \lesssim \delta y_{b} \lesssim -0.05$ as shown in the $R_{tb\tau}-\delta y_{b}$ plane. Also perfect YU solutions ($R_{tb\tau}=1$) can be realized in the range of  $-0.15 \lesssim \delta y_{b} \lesssim -0.09$. Such large negative contributions require large $\mu$-term when the NH terms are absent, but as displayed in the $\delta y_{b}-\mu$ plane, the YU solutions (red points) can be realized even when $\mu \gtrsim 1$ TeV. The YU impact on $A_{0}^{\prime}$ can also be seen in the $\delta y_{b}-A_{b}^{\prime}/m_{b}$ plane. The NH contributions to $y_{b}$ dominantly arise from $A_{0}^{\prime}$. When $A_{0}^{\prime} \sim 0$, $\delta y_{b}^{NH}$ is also negligible, and there are no YU solutions in these cases. In this context, negative threshold corrections happen with $A_{b}^{\prime} < 0$, and  its magnitude should be as large as about the sbottom mass. The $\delta yb^{H}-\delta y_{b}^{NH}$ plane also confirms that the holomorphic terms can vary freely, and they can even provide positive threshold corrections as large as about 0.3. Even such positive contributions can be compensated with larger negative NH contributions ($\delta y_{b}^{NH} \simeq -0.5$) and the solutions can lead to YU at $\mgut$.

\subsection{Mass Spectrum Compatible with YU}
\label{subsec:Mspec}

\begin{figure}[h!]
\centering
\includegraphics[scale=0.4]{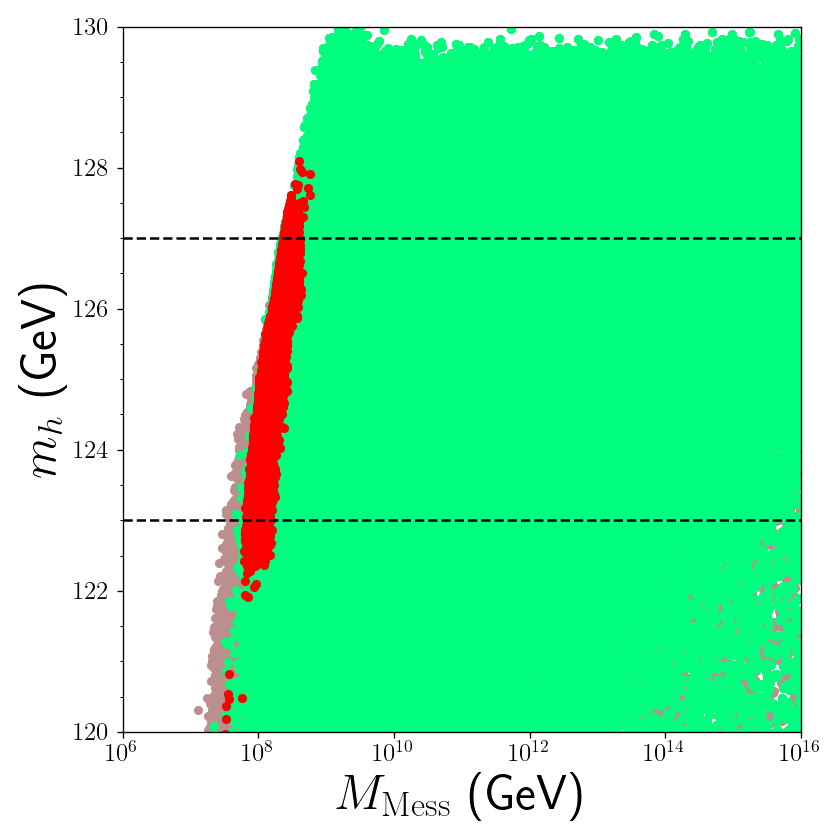}%
\includegraphics[scale=0.4]{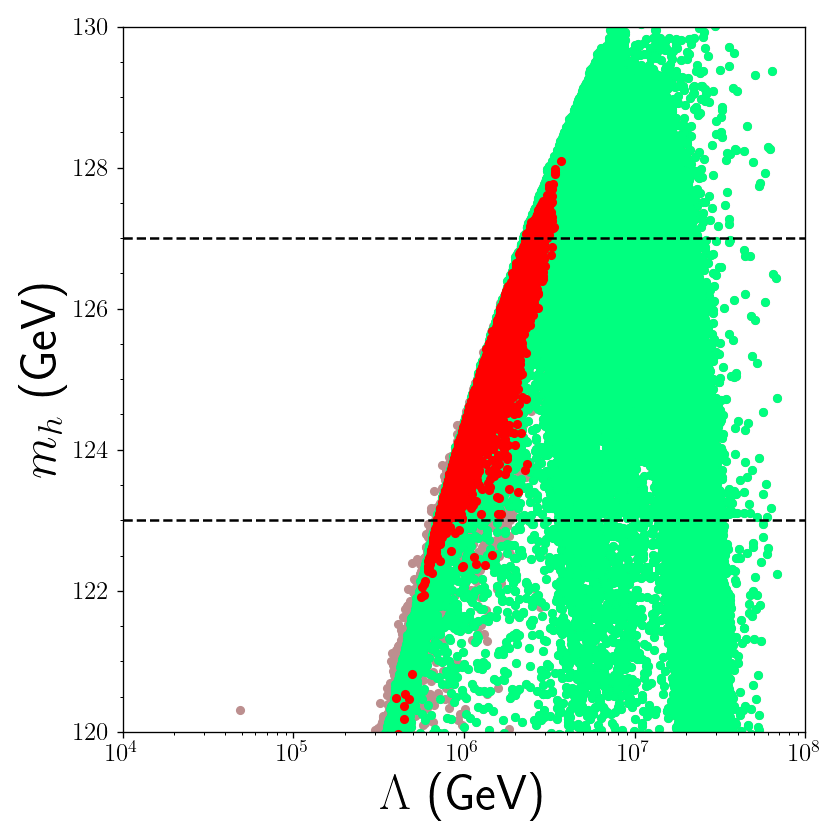}
\includegraphics[scale=0.4]{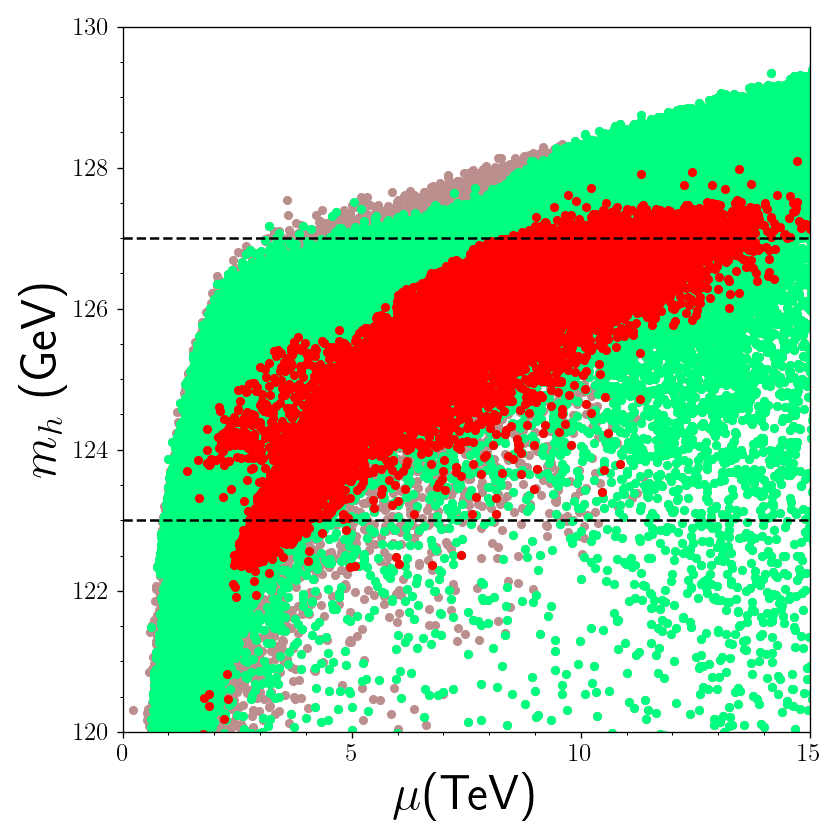}%
\includegraphics[scale=0.4]{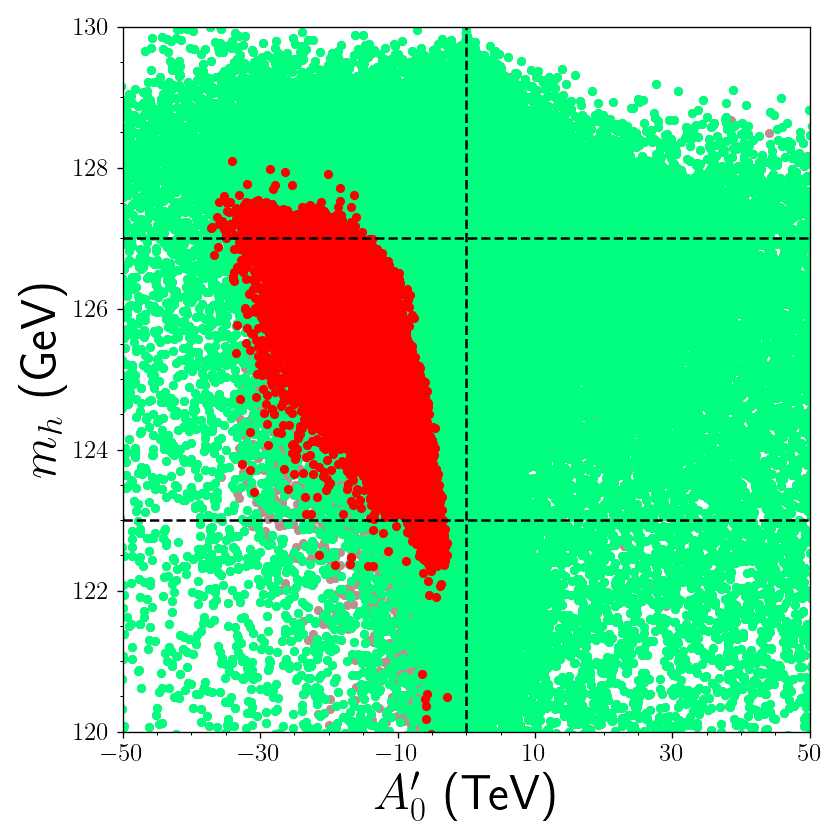}
\caption{The Higgs boson mass in the $m_{h}-M_{{\rm Mess}}$, $m_{h}-\Lambda$, $m_{h}-\mu$ and $m_{h}-A_{0}^{\prime}$ planes. The color coding is the same as in the bottom-right plane of Figure \ref{fig:YUHfund}. The red points form a subset of green and they display the YU solutions. The Higgs boson mass constraint is not applied in these planes, while its mass bounds are represented with the horizontal dashed lines.}
\label{fig:higgs} 
\end{figure}

The NH terms provide significant contributions to the Higgs boson mass as shown in Figure \ref{fig:higgs} in the $m_{h}-M_{{\rm Mess}}$, $m_{h}-\Lambda$, $m_{h}-\mu$ and $m_{h}-A_{0}^{\prime}$ planes. The color coding is the same as in the bottom-right plane of Figure \ref{fig:YUHfund}. The red points form a subset of green and they display the YU solutions. The Higgs boson mass constraint is not applied in these planes, while its mass bounds are represented with the horizontal dashed lines. In the absence of NH terms, the Higgs boson can barely weigh about 126 GeV, and the consistent mass scales for the Higgs boson can be realized only when the SUSY particles are quite heavy \cite{Ajaib:2012vc}. However, as shown in the top panels of Figure \ref{fig:higgs}, one can accommodate the Higgs boson in the spectrum even as heavy as about 128 GeV. As is well known, MSSM can lead to the consistent Higgs boson mass only when it can utilize large radiative corrections to its mass, which also require a large $\mu-$term in GMSB models. However, the $m_{h}-\mu$ plane shows that a Higgs boson mass of about 125 GeV can be realized even when $\mu \gtrsim 1$ TeV. In such solutions, the required radiative contributions to the Higgs boson mass can arise from the $A_{0}^{\prime}$ term. The $m_{h}-A_{0}^{\prime}$ plane shows that the a negative larger $A_{0}^{\prime}$ drives the Higgs boson to heavier mass scales in the YU compatible region (red points).

\begin{figure}[h!]
\centering
\includegraphics[scale=0.4]{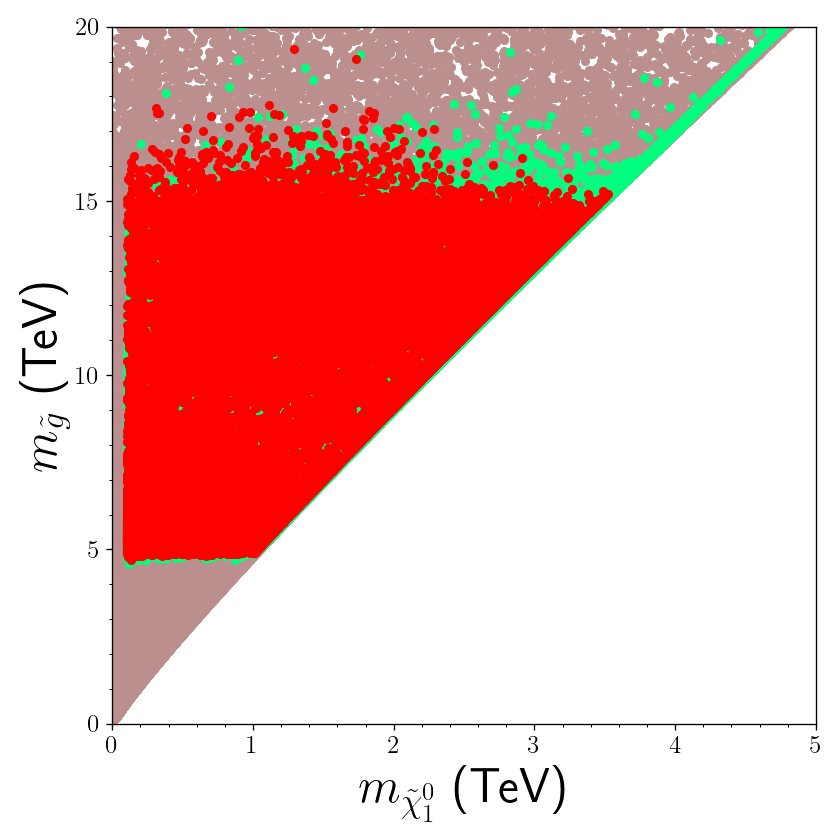}%
\includegraphics[scale=0.4]{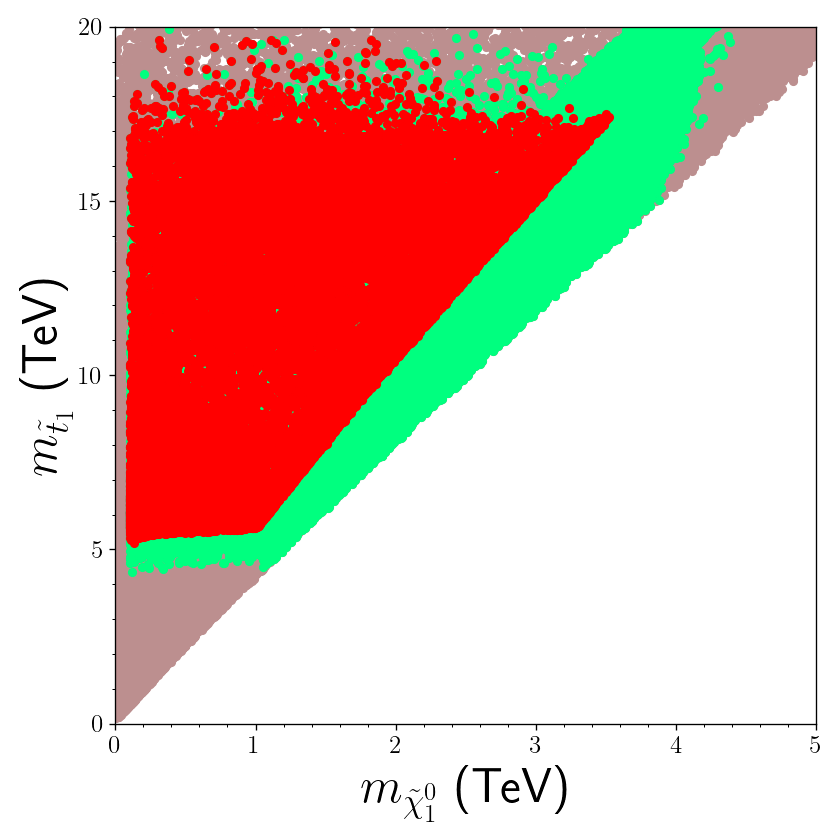}
\includegraphics[scale=0.4]{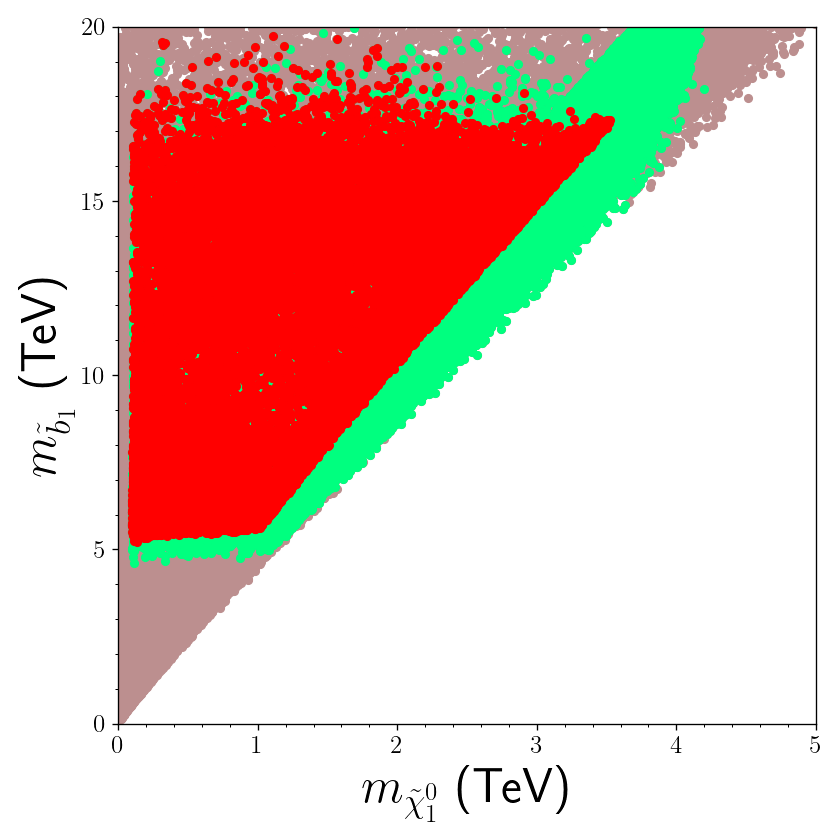}%
\includegraphics[scale=0.4]{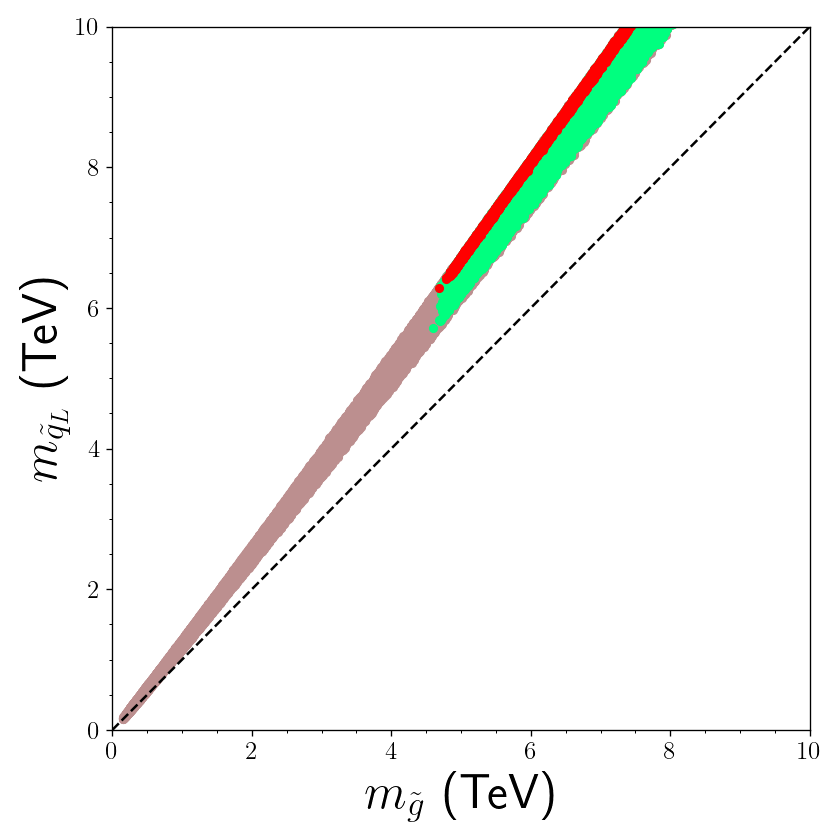}
\caption{The masses of strongly interacting SUSY particles in the $m_{\tilde{g}}-m_{\tilde{\chi}_{1}^{0}}$, $m_{\tilde{t}_{1}}-m_{\tilde{\chi}_{1}^{0}}$, $m_{\tilde{b}_{1}}-m_{\tilde{\chi}_{1}^{0}}$ and $m_{\tilde{q}_{L}}-m_{\tilde{g}}$ planes. The color coding is the same as in the bottom planes of Figure \ref{fig:YUHfund}. The diagonal line in the $m_{\tilde{q}_{L}}-m_{\tilde{g}}$ plane indicates the mass degeneracy between the first two-family squarks and gluino.}
\label{fig:SUSYQCD}
\end{figure}

\begin{figure}[h!]
\centering
\includegraphics[scale=0.4]{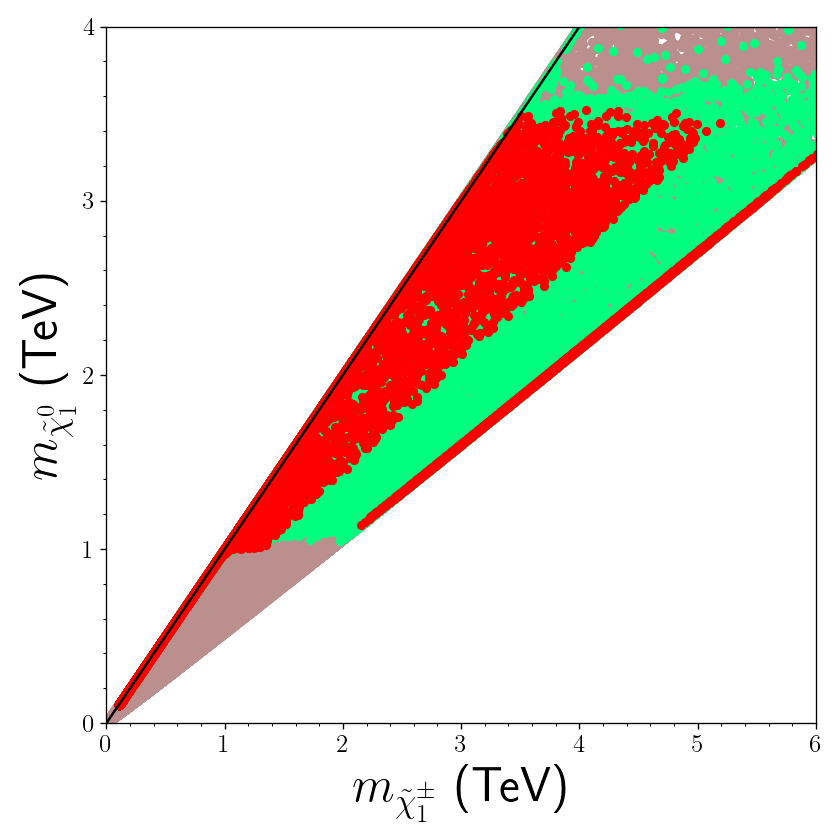}%
\includegraphics[scale=0.4]{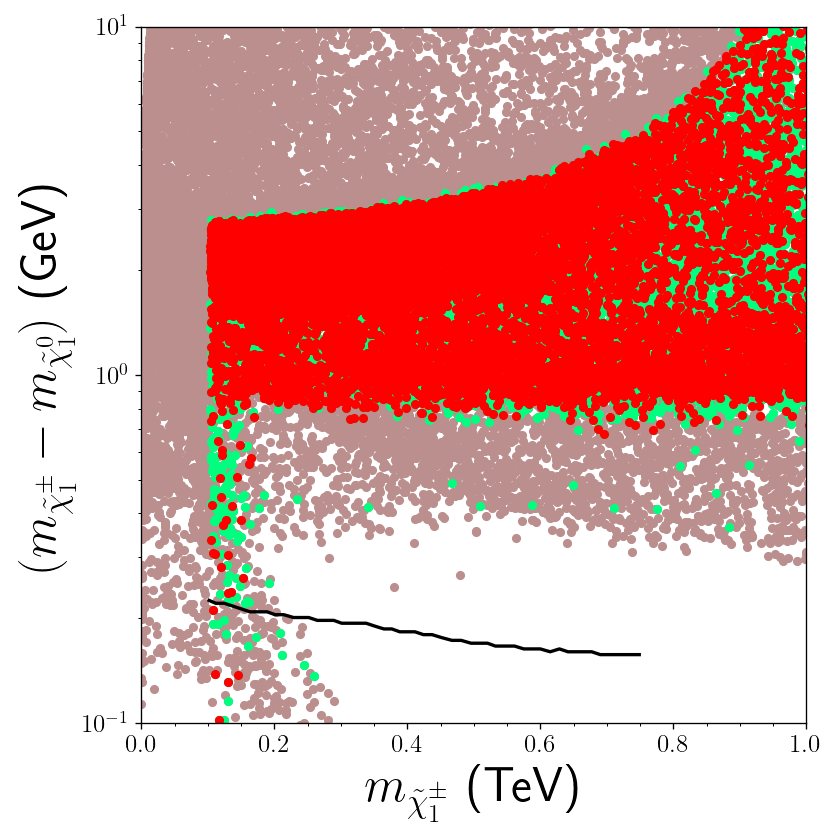}
\includegraphics[scale=0.4]{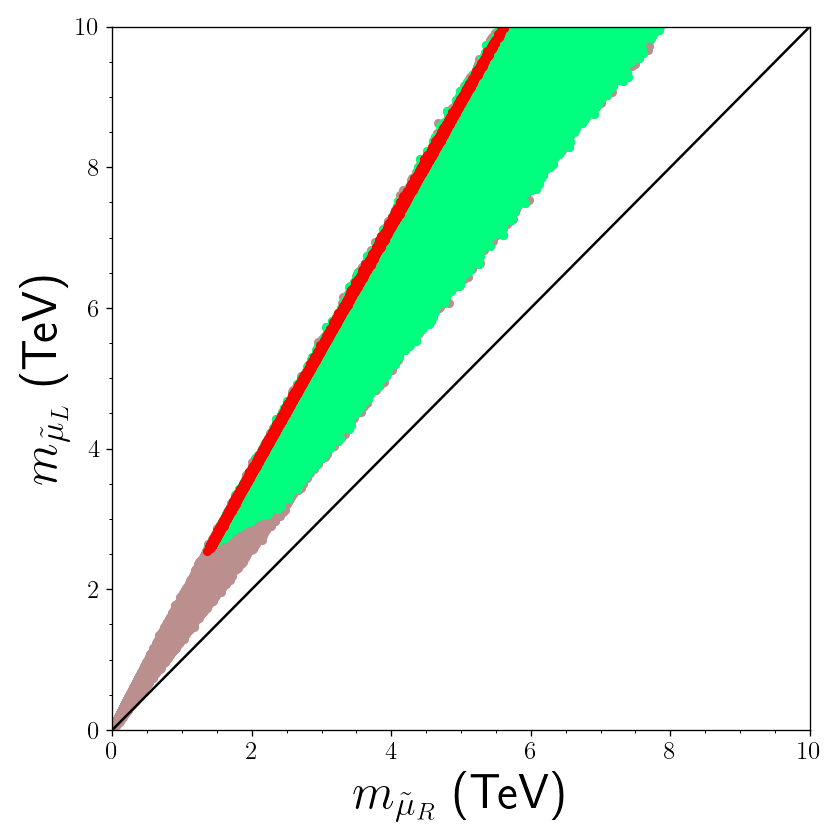}%
\includegraphics[scale=0.4]{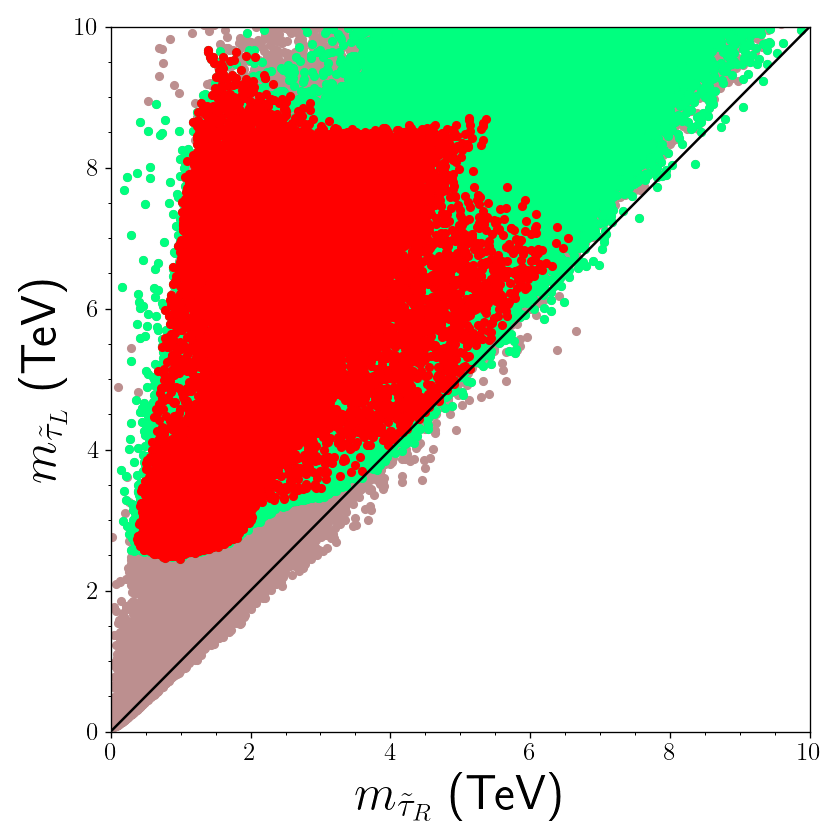}
\caption{The masses of electroweakinos (top) and sleptons (bottom) in the $m_{\tilde{\chi}_{1}^{0}}-m_{\tilde{\chi}_{1}^{\pm}}$, $(m_{\tilde{\chi}_{1}^{\pm}}-m_{\tilde{\chi}_{1}^{0}}) - m_{\tilde{\chi}_{1}^{\pm}}$, $m_{\tilde{\mu}_{L}} - m_{\tilde{\mu}_{R}}$ and $m_{\tilde{\tau}_{L}}-m_{\tilde{\tau}_{R}}$ planes. The color coding is the same as in the bottom planes of Figure \ref{fig:YUHfund}. The diagonal lines depict the mass degeneracy between the plotted particles. The black curve in the top-right plane represents the experimental results on the compressed spectra which exclude the solutions beneath it \cite{CMS:2023mny}.}
\label{fig:EWinos}
\end{figure}

The ranges of the fundamental parameters compatible with YU shown in Figure \ref{fig:YUHfund} typically lead to heavy spectrum for strongly interacting SUSY particles. Figure \ref{fig:SUSYQCD} displays their masses in the $m_{\tilde{g}}-m_{\tilde{\chi}_{1}^{0}}$, $m_{\tilde{t}_{1}}-m_{\tilde{\chi}_{1}^{0}}$, $m_{\tilde{b}_{1}}-m_{\tilde{\chi}_{1}^{0}}$ and $m_{\tilde{q}_{L}}-m_{\tilde{g}}$ planes. The color coding is the same as in the bottom planes of Figure \ref{fig:YUHfund}. The diagonal line in the $m_{\tilde{q}_{L}}-m_{\tilde{g}}$ plane indicates the mass degeneracy between the first two-family squarks and gluino. The top panels and bottom-left panel show that the stop, sbottom and gluino can only be as light as about 5 TeV when the solutions are compatible with the YU condition (red). In addition, the spectra involve the squarks of the first two families slightly heavier ($m_{\tilde{q}_{L}}\gtrsim 6$ TeV), as shown in the bottom-right panel. These mass scales are obviously beyond the sensitivity of collider experiments conducted at LHC, and they can even go beyond the High Luminosity LHC (HL-LHC) experiments. Our results for the mass spectrum of these particles can potentially be probed in FCC experiments \cite{FCC:2025lpp,FCC:2025uan,FCC:2025jtd}.

Despite the heavy masses of the strongly interacting SUSY particles, Figure \ref{fig:SUSYQCD} also reveals that, the neutralinos can be significantly light with $m_{\tilde{\chi}_{1}^{0}} \sim \mathcal{O}(100)$ GeV. This observation can lead to testable implications in the current experiments within the mGMSB framework. In this context, the plots in Figure \ref{fig:EWinos} show the masses of the electroweakly interacting particles in the $m_{\tilde{\chi}_{1}^{0}}-m_{\tilde{\chi}_{1}^{\pm}}$, $(m_{\tilde{\chi}_{1}^{\pm}}-m_{\tilde{\chi}_{1}^{0}}) - m_{\tilde{\chi}_{1}^{\pm}}$, $m_{\tilde{\mu}_{L}} - m_{\tilde{\mu}_{R}}$ and $m_{\tilde{\tau}_{L}}-m_{\tilde{\tau}_{R}}$ planes. The color coding is the same as in the bottom planes of Figure \ref{fig:YUHfund}. The diagonal lines depict the mass degeneracy between the plotted particles. The $m_{\tilde{\chi}_{1}^{0}}-m_{\tilde{\chi}_{1}^{\pm}}$ plane shows that the lightest chargino and neutralino can weigh from about 100 GeV to about 3.5 TeV, and they happen to be nearly degenerate when their masses are lighter than about 1 TeV. These solutions can be tested in the collider analyses over compressed spectra \cite{CMS:2023mny} whose results are represented with the black curve in the $(m_{\tilde{\chi}_{1}^{\pm}}-m_{\tilde{\chi}_{1}^{0}}) - m_{\tilde{\chi}_{1}^{\pm}}$ plane. In the region where the chargino and neutralino are nearly degenerate in mass, the mass difference can be as low as about $0.1$ GeV, and these solutions can be excluded by the current experimental bounds. Even though they are only a few in our scans, there are also solutions with $(m_{\tilde{\chi}_{1}^{\pm}}-m_{\tilde{\chi}_{1}^{0}}) \lesssim 1$ GeV are more likely to be tested in the next update from such analyses.

\begin{figure}[h!]
\centering
\includegraphics[scale=0.4]{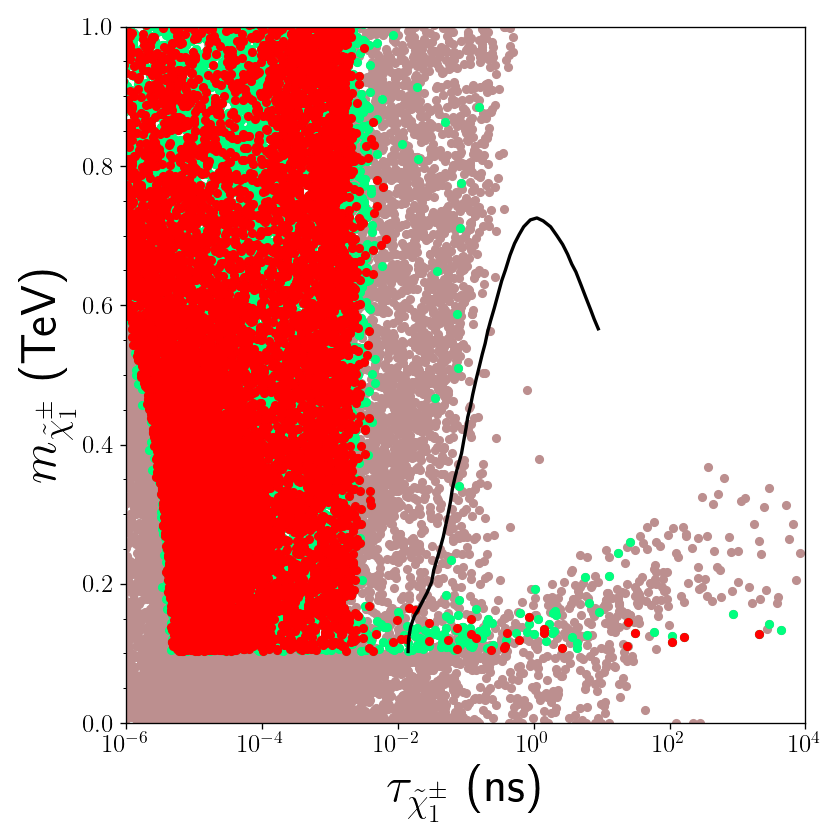}%
\includegraphics[scale=0.4]{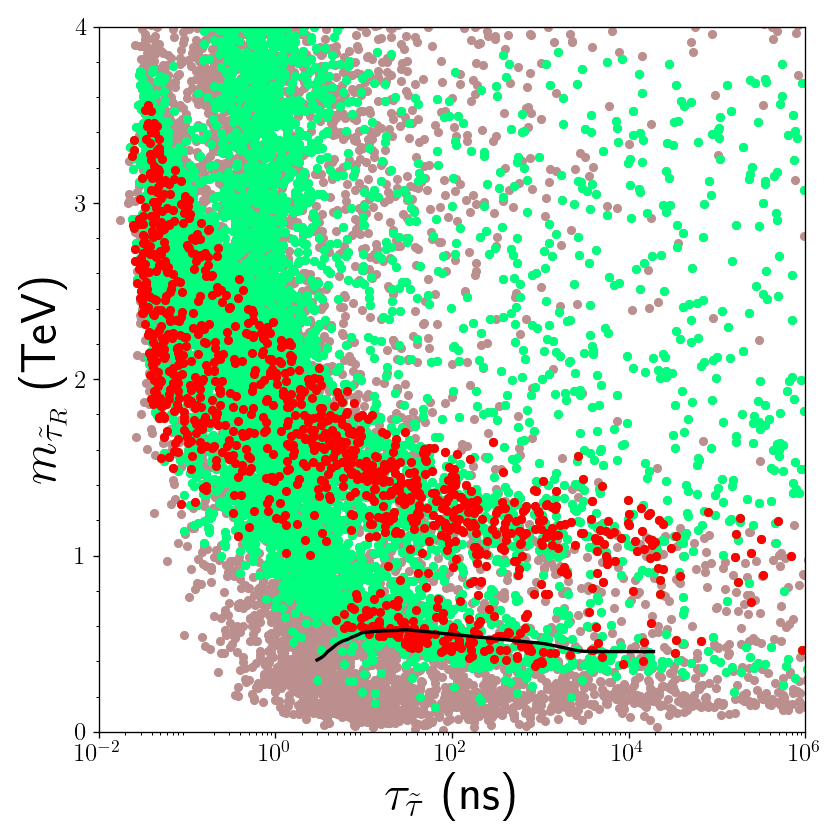}
\caption{The lifetimes of the lightest chargino (left) and stau (right). The color coding is the same as that described for the bottom planes of Figure \ref{fig:YUHfund}. The curves represent the results from the current analyses for chargino (left) and stau (right) lifetimes \cite{ATLAS:2025fdm,ATLAS:2026hnb}.}
\label{fig:LTCS}
\end{figure}

We also display our results for the slepton masses in the bottom panels of Figure \ref{fig:EWinos}. The sleptons from the first two families happen to be quite degenerate in their masses, and the bottom-left panel shows their mass scales shown in the smuon masses. The results in this panel shows that smuons (and selectrons since they are degenerate in mass) can only be as light as about $1.5-2$ TeV. On the other hand, the third family sleptons - staus - can be much lighter. While the left-handed stau is heavier than 2 TeV in the parameter space, the right-handed stau can be as light as about 100 GeV. The solutions with $m_{\tilde{\tau}_{R}}\lesssim 100$ GeV are excluded by the LEP bounds \cite{LEPWorkingGroupforHiggsbosonsearches:2003ing}.

Such light charged particles can also be subjected to the analyses for the long-lived particles. Smuons are typically known to be long-lived particles in the GMSB models; however, the current results \cite{ATLAS:2023iip} can probe such smuons only up to about 600 GeV. Therefore our results need to wait for the future experiments to probe YU through long-lived smuons. On the other hand, the observed mass scales for the chargino and stau can be within reach of the current and near future collider experiments \cite{ATLAS:2025fdm,ATLAS:2026hnb}. Figure \ref{fig:LTCS} shows the lifetimes of these charged particles in correlation with their masses. The analyses for the chargino can hold when the mass difference between the chargino and neutralino is close to the pion mass, and they can exclude some solutions in mGMSB models as shown in the $m_{\tilde{\chi}_{1}^{\pm}}-\tau_{\tilde{\chi}_{1}^{\pm}}$ plane. The possible longest lifetime for chargino in our results is realized to be about $10^{-2}$ ns (red points in the left side of the curve). In this region, the chargino can be even lighter than 200 GeV, and these solutions are expected to be tested as the sensitivity in these analyses is being improved. On the other hand, the staus can fly for about $10^{7}$ ns, and the solutions in which the staus can fly longer than about $10^{3}$ ns are already excluded by the analyses \cite{ATLAS:2025fdm}. The solutions with lower stau lifetime can be probed up to about 600 GeV. The $m_{\tilde{\tau}_{R}}-\tau_{\tilde{\tau}}$ plane displays many solutions which can receive impacts from the lifetime analyses of the stau. These analyses have already excluded some of such solutions, while mGMSB models can still yield many more to be tested in these analyses.

\subsection{Muon $g-2$ with the NH contributions}
\label{subsec:muong2NH}

\begin{figure}[h!]
\centering
\includegraphics[scale=0.4]{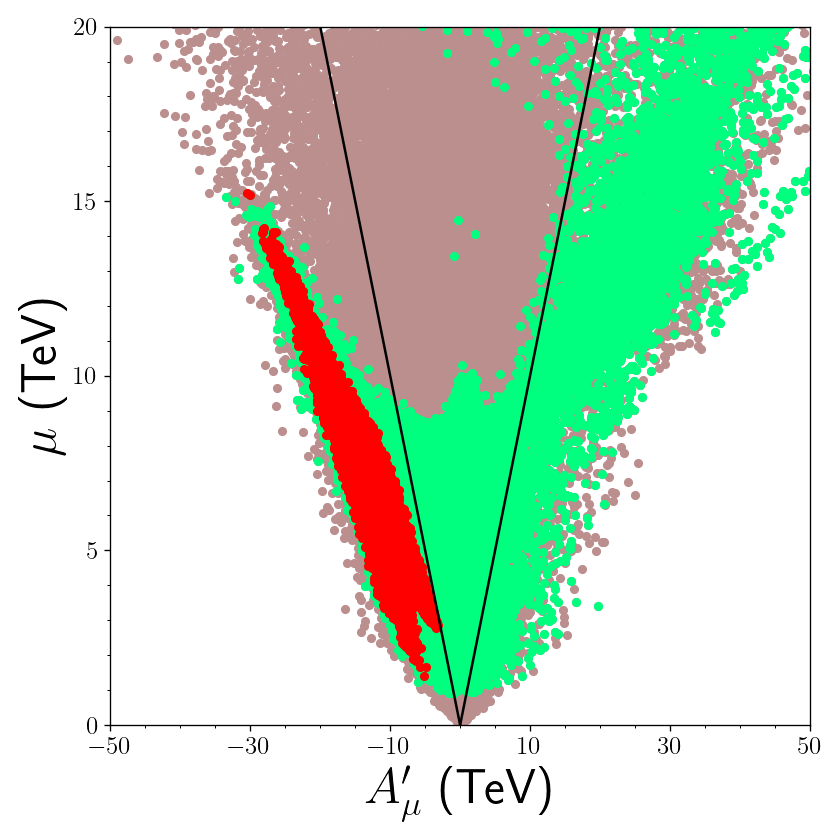}%
\includegraphics[scale=0.4]{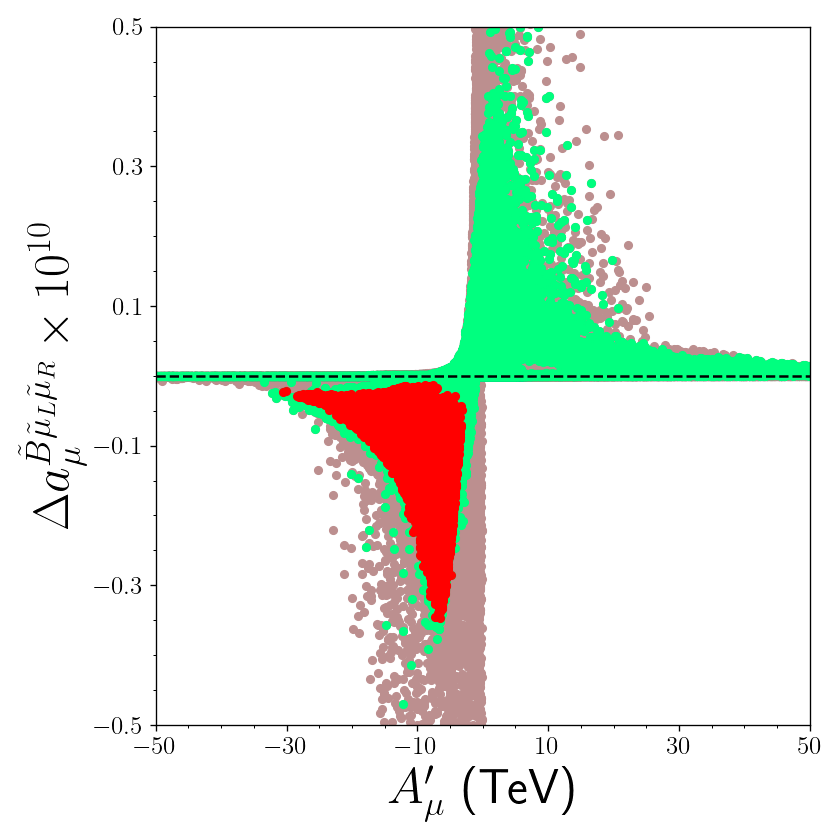}
\includegraphics[scale=0.4]{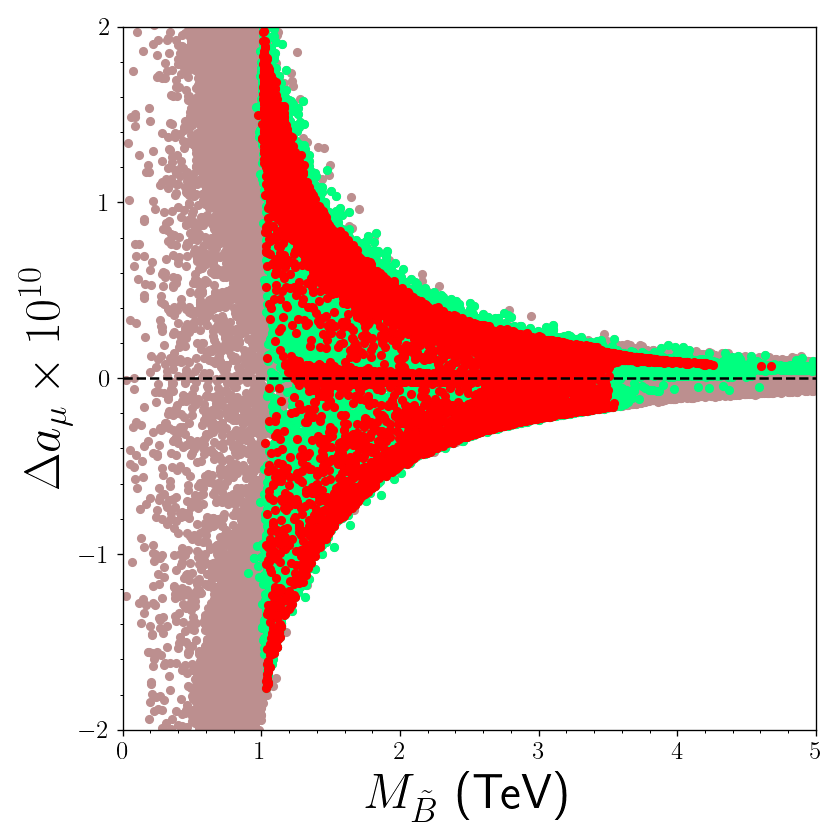}
\caption{The plots in the $\mu-A_{\mu}^{\prime}$, $\Delta a_{\mu}^{\tilde{B}\tilde{\mu}_{L}\tilde{\mu}_{R}}-A_{\mu}^{\prime}$ and $\Delta a_{\mu}-M_{\tilde{B}}$ planes. The color coding is the same as in the bottom planes of Figure \ref{fig:YUHfund}. The horizontal dashed line in the top-right panel depicts the solutions in which the muon $g-2$ contributions through the bino-smuon loop is zero. The diagonal lines in the top-left plane represent the regions with $\mu = \lvert A_{\mu}^{\prime}\rvert$. }
\label{fig:bmumudamu}
\end{figure}

As discussed in Section \ref{subsec:FPSYU}, the YU solutions favor negative $A_{0}^{\prime}$ values. In this context, the results also yield a negative $A_{\mu}^{\prime}$ which receives a restriction due to our constraint from the current muon $g-2$ measurements, which strictly forbids negative muon $g-2$ results in the YU compatible spectra within the minimal GMSB framework. As realized in most cases, the dominant supersymmetric contributions to muon $g-2$ arise from the bino-smuon loop which is represented with the top-left diagram in Figure \ref{fig:muong2diags}. The plots in Figure \ref{fig:bmumudamu} display our results relevant to these contributions with the plots in the $\mu-A_{\mu}^{\prime}$, $\Delta a_{\mu}^{\tilde{B}\tilde{\mu}_{L}\tilde{\mu}_{R}}-A_{\mu}^{\prime}$ and $\Delta a_{\mu}-M_{\tilde{B}}$ planes. The color coding is the same as in the bottom planes of Figure \ref{fig:YUHfund}. The horizontal dashed line in the top-right panel depicts the solutions in which the muon $g-2$ contributions through bino-smuon loop are zero. The diagonal lines in the top-left plane represent the regions with $\mu = \lvert A_{\mu}^{\prime}\rvert$.

An interesting results directly emerges from the $\mu-A_{\mu}^{\prime}$ plane that all the YU solutions (red) are accumulated in the left side of the left diagonal line in this plane. These solutions turn the factor $\mu + A_{\mu}^{\prime}$ negative, and consequently, all the supersymmetric contributions through the bino-smuon loop become negative even in the solutions which are shown around the left diagonal line, as seen in the $\Delta a_{\mu}^{\tilde{B}\tilde{\mu}_{L}\tilde{\mu}_{R}}-A_{\mu}^{\prime}$ plane. The bino-smuon loop can contribute to muon $g-2$ as large as about $-0.3$, and the results show that one needs mostly a positive $A_{\mu}^{\prime}$ (green above the horizontal dashed line) to realize positive muon $g-2$ contributions through the bino-smuon loop, but this region is not compatible with the YU condition. Note that, the overall muon $g-2$ constraint is applied the results in the top panels. Therefore, the green and red solutions mean that the overall muon $g-2$ contributions can still be positive despite the negative contributions through the bino-smuon loop. The bottom panel shows our results for the overall muon $g-2$ contributions, and indeed, one can still realize positive contributions as large as about $1.5\times 10^{-10}$, and in this region bino mass ($M_{\tilde{B}}$) cannot be smaller than about 1 TeV.

In this context, one can arrive at the conclusion that the SUSY spectra should be tuned such that the main supersymmetric contributions do not originate to the bino-smuon loop. As discussed in Section \ref{subsec:muong2NHd}, the contributions approximated by Eqs.(\ref{eq:whcont} and \ref{eq:charcount}) are also negative due to the negative factor $\mu + A_{\mu}^{\prime}$. In this case, the only option that can suppress these negative contributions and turn the overall muon $g-2$ contributions positive is the right-handed smuon-bino-higgsino loop, which can be calculated with the second loop factor in the parentheses of Eq.(\ref{eq:bhcont}). Note that, the left-handed smuons also contribute through this diagram (the first loop factor in the parentheses) and its contributions are also expected to be negative. However, as shown in Figure \ref{fig:EWinos}, left-handed smuons are mostly about twice as heavy as the right-handed smuons, the contributions involving the left-handed smuons are more suppressed compared to those from the right-handed smuons.

\begin{figure}[h!]
\centering
\includegraphics[scale=0.4]{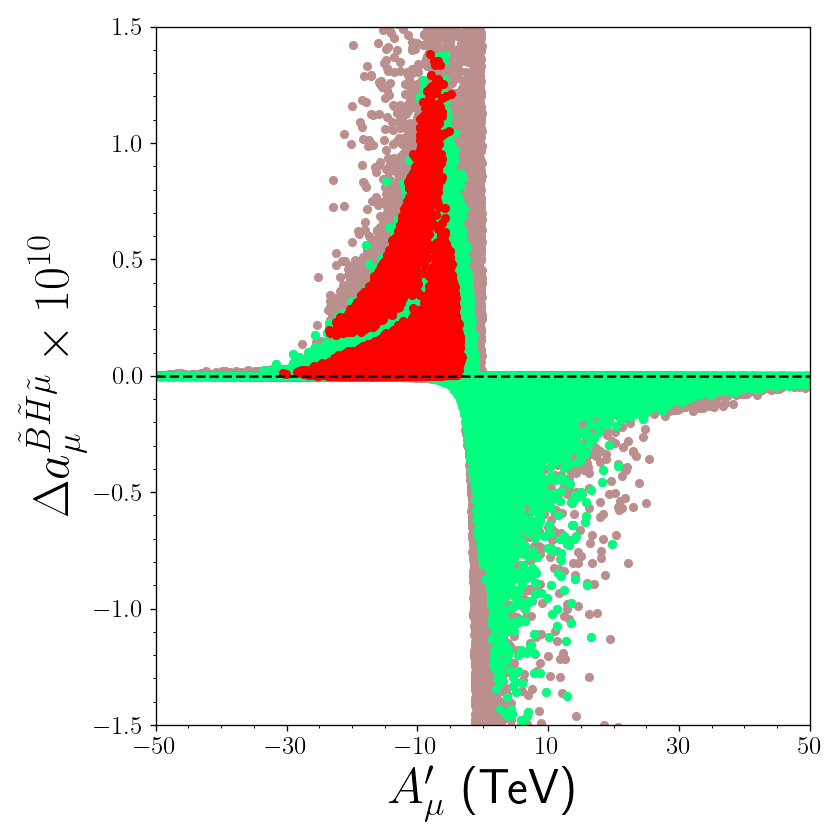}%
\includegraphics[scale=0.4]{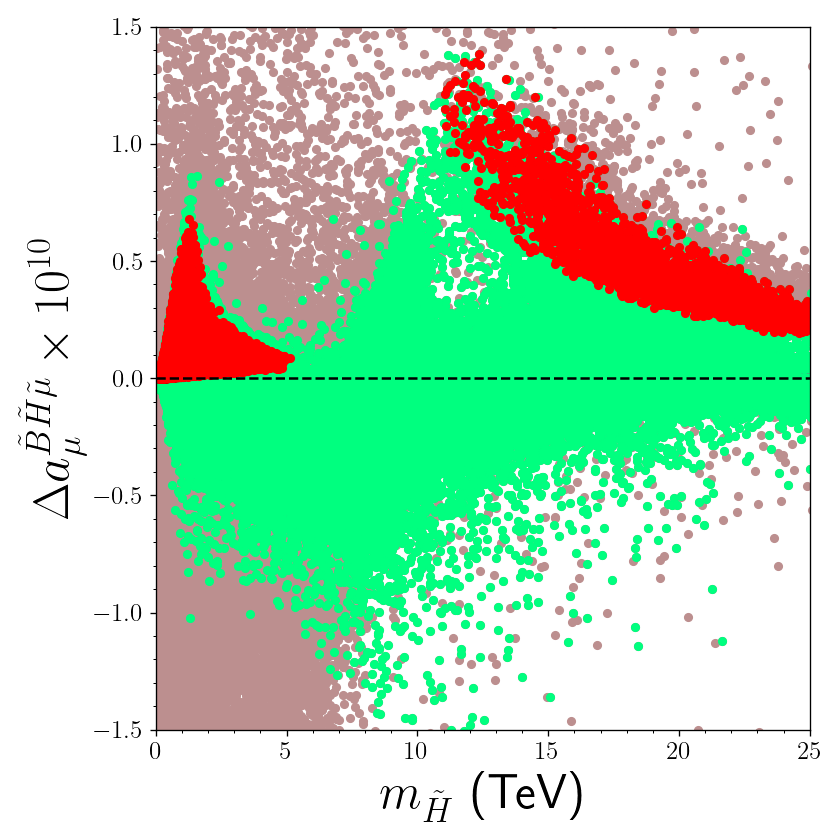}
\includegraphics[scale=0.4]{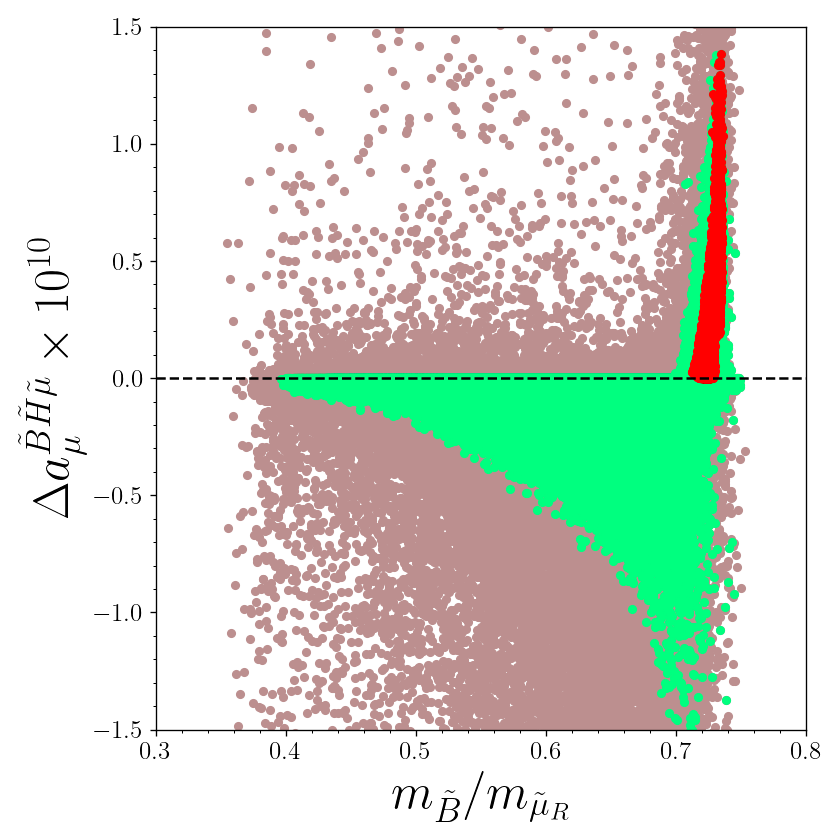}%
\includegraphics[scale=0.38]{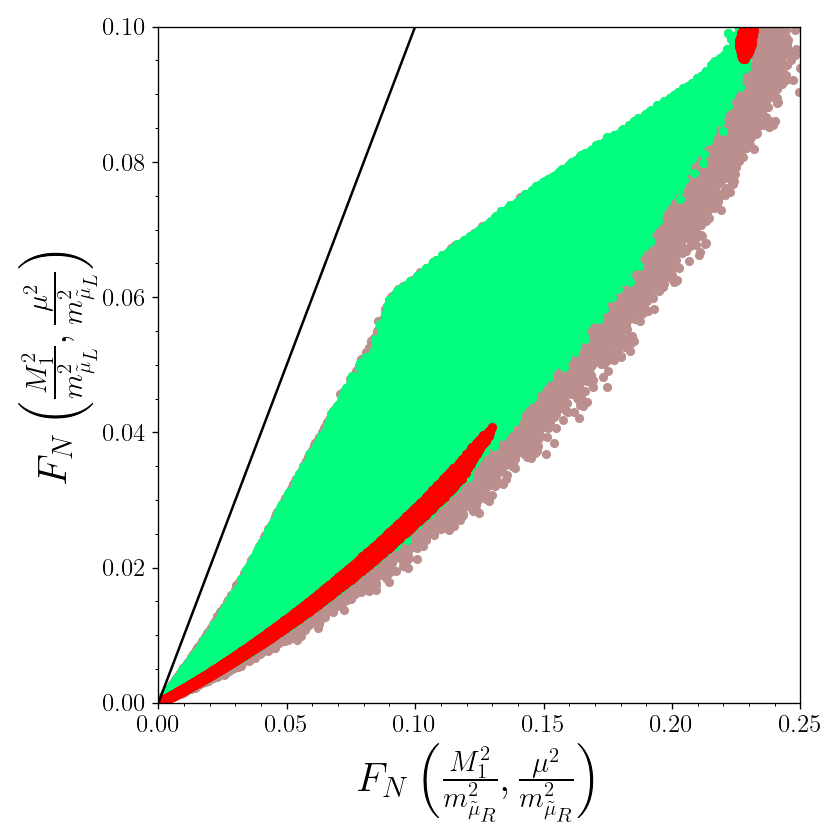}
\caption{The supersymmetric muon $g-2$ contributions through the bino-higgsino-smuon loop in correlation with $A_{\mu}^{\prime}$, the higgsino mass ($m_{\tilde{H}} \equiv \lvert \mu + \mu^{\prime}\rvert$ (top panels), and bino ($M_{\tilde{B}}$) and right-handed smuon masses (bottom-left panel). We also show the behaviour of the loop factors in the bottom-right panel. The color coding is the same as in Figure \ref{fig:bmumudamu}. The horizontal dashed lines depict the solutions in which the muon $g-2$ contributions from bino-higgsino-smuon loop are zero. The diagonal line in the bottom-right panel shows the regions where the loop factors are equal to each other and make the relevant contributions zero.}
\label{fig:bhsloop}
\end{figure}

Figure \ref{fig:bhsloop} displays our results for the contributions from the smuon-bino-higgsino loop in correlation with $A_{\mu}^{\prime}$, the higgsino mass ($m_{\tilde{H}} \equiv \lvert \mu + \mu^{\prime}\rvert$ (top panels), and bino ($M_{\tilde{B}}$) and right-handed smuon masses (bottom-left panel). We also show the behaviour of the loop factors in the bottom-right plane. The color coding is the same as in Figure \ref{fig:bmumudamu}. The horizontal dashed lines depict the solutions in which the muon $g-2$ contributions from the bino-higgsino-smuon loop are zero. The diagonal line in the bottom-right plane shows the regions where the loop factors are equal to each other and make the relevant contributions zero. Note that these contributions involve those from both the left-handed and right-handed smuons. As seen from the $\Delta a_{\mu}^{\tilde{B}\tilde{H}\tilde{H}}-A_{\mu}^{\prime}$ plane, a negative $A_{\mu}^{\prime}$ yields positive contributions, and they can be as large as about $1.5\times 10^{-10}$. Another interesting result can be observed in the  $\Delta a_{\mu}^{\tilde{B}\tilde{H}\tilde{H}}-m_{\tilde{H}}$ plane. Sufficiently large and positive contributions can be realized in two distinct regions. In the first region, the Higgsino mass is lighter than about 5 TeV, and the maximum contribution to muon $g-2$ reaches about $0.7\times 10^{-10}$ when $m_{\tilde{H}} \lesssim 2$ TeV. The results in this region can partially explain the overall positive muon $g-2$ results in our analyses. Interestingly, the largest muon $g-2$ contributions through the bino-higgsino-smuon loop are realized in the second region where the higgsino mass cannot be lighter than about 10 TeV. In the regions where $m_{\tilde{H}}\gg M_{\tilde{B}},m_{\tilde{\mu}}$, the spectrum becomes highly fine-tuned, and in the limit of heavy Higgsinos, the loop functions given in Eqs.(\ref{eq:loopN1} and \ref{eq:loopN2}) asymptotically behave as $F_{N}\simeq M_{\tilde{B}}/m_{\tilde{\mu}}$. The largest positive contribution in this region is about $1.5\times 10^{-10}$, and it can drive the overall muon $g-2$ contributions to positive values. These values can be realized when $M_{\tilde{B}}/m_{\tilde{\mu}_{R}}\simeq 0.7$ as shown in the bottom-left panel of Figure \ref{fig:bhsloop}.

\begin{table}[h!]
\centering
\setstretch{1.8}\scalebox{0.75}{\begin{tabular}{|c|ccccc|}
\hline  & Point 1 & Point 2 & Point 3 & Point 4 & Point 5 \\ \hline
$\Lambda$ &  $ 1.5 \times 10^{6} $  &  $ 1.6 \times 10^{6} $  &  $ 1.5 \times 10^{6} $  &  $ 1.8 \times 10^{6} $  &  $ 1.1 \times 10^{6} $  \\
$M_{{\rm Mess}}$ &  $ 1.7 \times 10^{8} $  &  $ 1.1 \times 10^{9} $  &  $ 7.6 \times 10^{8} $  &  $ 1.8 \times 10^{8} $  &  $ 1.3 \times 10^{8} $  \\
$\tan\beta$ &  $ 52.7 $  &  $ 57.2 $  &  $ 56.1 $   &  $ 50.4 $  &  $ 51.6 $  \\
$A_{0}^{\prime}$ &  $ -13609 $  &  $ \cred{-2288} $  &  $ -3040 $  &  $ -18605 $  &  $ -10136 $  \\
$\mu^{\prime}$ &  $ -7142 $  &  $ -13708 $  &  $ -13899 $  &  $ -8320 $  &  $ -4520 $  \\ \hline
$\mu$ &  $ 6707 $  &  $ 2100 $  &  $ \cred{1849} $  &  $ 8465 $  &  $ 5216 $  \\
$m_{h}$ &  $ \cred{125.6} $  &  $ \cred{125.6} $  &  $ \cred{125.6} $ &  $ \cred{125.9} $  &  $ 124.5 $  \\
$m_{H}$ &  $ 2355 $  &  $ 2265 $  &  $ 2030 $  &  $ 3080 $  &  $ 2010 $  \\
$m_{A}$ &  $ \cred{2355} $  &  $ \cred{2030} $  &  $ \cred{2265} $  &  $ 3080 $  &  $ \cred{2010} $  \\
$m_{H^{\pm}}$ &  $ 2356 $  &  $ 2264 $  &  $ 2029 $  &  $ 3080 $  &  $ 2012 $  \\ \hline
$m_{\tilde{\chi}_{1}^{0}}$,$m_{\tilde{\chi}_{2}^{0}}$ &  $ \cred{469.0} $ ,  $ \cred{471.5} $  &  $ 2181 $ ,  $ 4045 $  &  $ 2115 $ ,  $ 3927 $ &  $ \cred{120.3} $ ,  $ \cred{122.3} $  &  $ 697.2 $ ,  $ 700.8 $  \\
$m_{\tilde{\chi}_{3}^{0}}$,$m_{\tilde{\chi}_{4}^{0}}$ &  $ 2129 $ ,  $ 3968 $  &  $ 11592 $ ,  $ 11592 $  &  $ 12022 $ ,  $ 12022 $  &  $ 2568 $ ,  $ 4798 $  &  $ 1579 $ ,  $ 2967 $  \\
$m_{\tilde{\chi}_{1}^{\pm}}$,$m_{\tilde{\chi}_{2}^{\pm}}$ &  $ \cred{470.4} $ ,  $ 3968 $  &  $ 4045 $ ,  $ 11592 $  &  $ 3927 $ ,  $ 12022 $  &  $ \cred{120.9} $ ,  $ 4798 $  &  $ 699.0 $ ,  $ 2967 $  \\ \hline
$m_{\tilde{g}}$ &  $ 9617 $  &  $ 9832 $  &  $ 9564 $  &  $ 11379 $  &  $ 7335 $  \\
$m_{\tilde{u}_{1}}$,$m_{\tilde{u}_{2}}$ &  $ 12457 $ ,  $ 13393 $  &  $ 12453 $ ,  $ 13423 $  &  $ 12142 $ ,  $ 13076 $  &  $ 14808 $ ,  $ 15960 $  &  $ 9463 $ ,  $ 10155 $  \\
$m_{\tilde{t}_{1}}$,$m_{\tilde{t}_{2}}$ &  $ 10793 $ ,  $ 11894 $  &  $ 12983 $ ,  $ 13881 $  &  $ 12787 $ ,  $ 13642 $  &  $ 12575 $ ,  $ 13961 $  &  $ 8089 $ ,  $ 8931 $  \\ \hline
$m_{\tilde{d}_{1}}$,$m_{\tilde{d}_{2}}$ &  $ 12339 $ ,  $ 13394 $  &  $ 12311 $ ,  $ 13424 $  &  $ 12008 $ ,  $ 13076 $  &  $ 14666 $ ,  $ 15960 $  &  $ 9378 $ ,  $ 10156 $  \\
$m_{\tilde{b}_{1}}$,$m_{\tilde{b}_{2}}$ &  $ 10762 $ ,  $ 11894 $  &  $ 12786 $ ,  $ 13880 $  &  $ 12597 $ ,  $ 13642 $  &  $ 12566 $ ,  $ 13961 $  &  $ 8104 $ ,  $ 8930 $  \\ \hline
$m_{\tilde{\nu}_{e}}$,$m_{\tilde{\nu}_{\tau}}$ &  $ 5475 $ ,  $ 5204 $  &  $ 5657 $ ,  $ 6915 $  &  $ 5468 $ ,  $ 6724 $  &  $ 6613 $ ,  $ 6082 $  &  $ 4090 $ ,  $ 3784 $  \\
$m_{\tilde{e}_{1}}$,$m_{\tilde{e}_{2}}$ &  $ 3036 $ ,  $ 5476 $  &  $ 3205 $ ,  $ 5658 $  &  $ 3080 $ ,  $ 5469 $  &  $ 3664 $ ,  $ 6614 $  &  $ 2254 $ ,  $ 4091 $  \\
$m_{\tilde{\tau}_{1}}$,$m_{\tilde{\tau}_{2}}$ &  $ 1905 $ ,  $ 5206 $  &  $ 6466 $ ,  $ 6916 $  &  $ 6326 $ ,  $ 6726 $  &  $ 505.1 $ ,  $ 6084 $  &  $ \cred{603.5} $ ,  $ 3786 $  \\ \hline
$\tau_{\tilde{\tau}}$ &  $ 3.4 \times 10^{-17} $  &  $ 2.4 \times 10^{-17} $  &  $ 2.4 \times 10^{-17} $  &  $ 1.3 \times 10^{-16} $  &  $ \cred{1.3 \times 10^{1}} $  \\ 
$\tau_{\tilde{\chi}_{1}^{\pm}}$ &  $ 1.7 \times 10^{-4} $  &  $ 2.1 \times 10^{-12} $  &  $ 2.1 \times 10^{-12} $  &  $ \cred{1.3 \times 10^{-2}} $  &  $ 4.4 \times 10^{-15} $  \\ \hline
$\delta y_{b^{H}}$ &  $ 0.19 $  &  $ 0.05 $  &  $ 0.04 $  &  $ 0.19 $  &  $ 0.20 $  \\
$\delta y_{b^{NH}}$ &  $ -0.31 $  &  $ \cred{-0.10} $  &  $ \cred{-0.11} $  &  $ -0.35 $  &  $ -0.31 $  \\
$R_{tb\tau}$ &  $ \cred{1.00} $  &  $ 1.06 $  &  $ 1.03 $  &  $ \cred{1.01} $  &  $ 1.09 $  \\ \hline
\end{tabular}
}
\caption{The benchmark points exemplifying our findings. The points are selected as being consistent with all the constraints employed in our analyses. They are additionally required to depict the Higgs boson mass as close as to its experimentally observed mass. All the parameters with mass dimension are listed in GeV, and the lifetime of particles is given in ns. {The colored values highlight the emphasized features of the benchmark points in regards to SM-like Higgs boson, relatively light masses and lifetimes of SUSY particles, and YU implications.}}
\label{tab:BPs}
\end{table}

The results discussed above can be summarized by analyzing the loop functions, which depend on the left-handed (vertical axis) and right-handed (horizontal axis) smuon masses in the bottom-right panel. As seen from their possible ranges, the loop function involving the right-handed smuon yields values about 2.5 times as large as that involving the left-handed smuon. Note that the loop function represented on the vertical axis is suppressed further, because $m_{\tilde{\mu}_{L}} > m_{\tilde{\mu}_{R}}$ as shown in Figure \ref{fig:EWinos}. In the tuned region where $m_{\tilde{H}}\gg M_{\tilde{B}},m_{\tilde{\mu}}$, this loop function involving the right-handed smuon can be as large as about 0.23, and it yields sufficiently large and positive muon $g-2$ contributions.

Before concluding, we display five benchmark points exemplifying our findings. The points are selected to be consistent with all the constraints employed in our analyses. They are additionally required to depict the Higgs boson mass as close as possible to its experimentally observed value. All the parameters with mass dimension are listed in GeV, and the lifetimes of particles are given in ns. The colored values highlight the emphasized features of the benchmark points. Point 1 represents solutions which lead to the perfect YU. In these solutions, the mass difference between the chargino and neutralino is about 1 GeV, and they can be tested in the analyses of compressed spectra in near future. Point 2 depicts a solution for the minimal NH contributions to $y_{b}$ to be compatible with the YU condition. In this solution, the holomorphic contributions are realized to be about 0.05, and they cannot unify the Yukawa couplings without NH contributions. Point 3 displays the solutions for minimum ranges of $\mu-$term, which is about 1.8 TeV in our scans. Point 4 exemplifies the solutions leading to a light chargino and neutralino ($\sim 120$ GeV). These solutions yield a mass difference of about 0.6 GeV between the chargino and neutralino, and they are more likely to be tested by upcoming upgrades in the analyses over the compressed spectra. In addition, these solutions lead to chargino lifetime of about $1.3\times 10^{-2}$. Such a lifetime is close to the current bounds (at the left end of the exclusion curve) given in the left plane of Figure \ref{fig:LTCS}, and a slight improvement in the analyses can easily probe these solutions. Finally Point 5 displays solutions which can also be tested in the analyses for the long lived particles. In these solutions, the stau weighs about 600 GeV, and they have relatively long lifetime. These solutions are more likely to be tested through the stau-lifetime by probably next updates in the analyses. We also observe that the solutions represented with Point 5 yield relatively light SM-like Higgs boson, and Point 5 displays the heaviest possible SM-like Higgs boson for these solutions, which is about 1 GeV lighter than the experimentally observed mass, and it is considered acceptable in our analyses.

All the points in Table \ref{tab:BPs}, except Point 4, can also be subjected to the heavy Higgs searches. The current results from these analyses can exclude the heavy scalar states up to about 2 TeV, when $\tan\beta \gtrsim 50$ \cite{ATLAS:2020zms,CMS:2022goy}. These solutions involve heavy Higgs bosons whose mass scales are very close to the exclusion, and they can be tested in heavy Higgs searches when the next updates are revealed. Note that even though we display only the CP-odd Higgs boson mass in color, the experimental analyses can shape the whole Higgs spectrum.

{Our results can be constrained further and/or probed by the dark matter observations and calculations primarily focusing on gravitino LSP scenarios. Minimally constructed GMSB models typically accommodate a gravitino LSP, whose mass lies between the eV and MeV scales \cite{Giudice:1998bp,Ajaib:2012vc,Gogoladze:2015tfa}. Due to the weakness of its gravitational interactions, the gravitino dark matter can be thermalized in the early universe, and hence its abundance can be frozen out when it is still relativistic. In this case, the gravitino should be lighter than about 1 keV and the required dilution in its abundance can be realized at low reheating temperatures \cite{Moroi:1993mb,Roszkowski:2014lga}. However, in this case, the scenario becomes problematic for thermal leptogenesis, which requires $T_{{\rm Reh}}\gtrsim 10^{9}$ GeV \cite{Pradler:2006qh,Georgilas:2026bqf}. Another problem with a low reheating temperature is that the DM gravitino cannot be produced at the right relic density. \cite{Rychkov:2007uq}. Indeed, this might be desirable since within this mass scale, the gravitino can be considered to be warm DM, which is allowed to form the DM relic density only up to about $15\%$ \cite{Viel:2005qj,Feng:2010ij}. On the other hand, the DM implications for a gravitino LSP can be drastically different in non-standard scenarios which may assume the presence of axion-like particles \cite{Deshpande:2023zed} or entropy dilution \cite{Nemevsek:2022anh}, etc. These scenarios can be explored through more robust and comprehensive analyses which are left for a possible future work.}

\section{Conclusion}
\label{sec:conc}

We explore the low scale implications of YU in minimally constructed GMSB models in the presence of NH terms. Among these terms, we assume the higgsino mass receives contributions from $\mu^{\prime}$, which can be induced through several mechanisms, while the other NH terms are generated only through the SUSY breaking. These models are known to receive a strong and negative impact from the Higgs boson mass constraint. Previous studies show that they can accommodate consistent Higgs boson mass in the spectrum only when the SUSY particles are quite heavy. In addition, once the YU condition is imposed at $\mgut$, only the models with $\mu < 0$ can be compatible with the YU solutions, since the solutions with $\mu > 0$ cannot provide large negative threshold contributions to $y_{b}$, which are needed to realize YU solutions. On the other hand, our analyses show that the required negative contributions to $y_{b}$ can be provided by the NH terms, and YU solutions can also be realized in models of the GMSB class even when $\mu > 0$. 

These required negative contributions can arise from negative values of $A_{b}^{\prime}$. In this case, the globally negative sign of $A_{i}^{\prime}$ terms lead to negative $A_{\mu}^{\prime}$. Consequently, the parameter space compatible with YU typically overlaps with regions where the supersymmetric corrections to the muon (g−2) are found to be negative. However, our analyses show that the SUSY mass spectrum in the YU compatible region is tuned such that the muon $g-2$ contributions from the right-handed smuon-bino-higgsino loop are sufficiently large and positive which suppress the  negative contributions.

We also find that the SM-like Higgs boson mass can easily be accommodated consistently, but it still requires the strongly interacting SUSY particles to be heavy. The typical mass scales for these particles are found to be about 5 TeV or higher, which is beyond the sensitivity of current collider experiments. On the other hand, the electroweakly interacting SUSY particles do not have to be heavy, and such solutions can be subjected to the analyses over these particles. We observe that the chargino can be lighter than about 1 TeV, and it is nearly degenerate with the lightest neutralino in these solutions. The current experimental analyses can probe such solutions when the chargino is lighter than about 800 GeV and the mass difference between the chargino and the lightest neutralino is less than about 0.2 GeV. Solutions with $m_{\tilde{\chi}_{1}^{0}} \gtrsim 120$ GeV can be realized and they are expected to be tested soon in the analyses over the compressed SUSY spectra. They can also be probed by lifetime analyses of the chargino and stau. Testing such light charginos via their lifetimes is only a matter of sensitivity in the experimental analyses, and a slight improvement can easily probe them. In the regions with light charginos, we also find that the stau can be light as well. Our results show that staus of mass about 600 GeV can be consistent with the current analyses over its lifetime and they are more likely to be tested very soon in the analyses over the final states with disappearing tracks. We exemplify our findings with five benchmark points, which also reveal that such solutions can be probed soon in the heavy Higgs boson searches.

\noindent{\bf Acknowledgment}

The work of BN and CSU is supported in part by the Scientific and Technological Research Council of Turkey (TUBITAK) Grant No. MFAG-125F122. The numerical calculations reported in this paper were partially performed at TUBITAK ULAKBIM, High Performance and Grid Computing Center (TRUBA resources).

\providecommand{\href}[2]{#2}\begingroup\raggedright\endgroup


\end{document}